\documentclass[journal,twoside]{IEEEtran}
\usepackage{amsmath,amssymb,amsfonts}
\usepackage{graphicx}
\usepackage[caption=false,font=footnotesize]{subfig}
\usepackage{hyperref}
\usepackage{cite}
\usepackage{booktabs}

\begin{document}
\bstctlcite{IEEEexample:BSTcontrol}

\title{Extreme Value Theory for Peak Power in Discrete-Time Complex Gaussian Processes}

\author{Adam Wunderlich
\thanks{The author is with the Communications Technology Laboratory, National Institute of Standards and Technology, Boulder, Colorado, 80305 USA. \newline (e-mail: adam.wunderlich@nist.gov).}}


\IEEEpubid{U.S. Government work not protected by U.S. copyright.}

\maketitle

\begin{abstract}
The peak power for a band-limited, radio frequency (RF) signal, often expressed using the dimensionless peak-to-average power ratio (PAPR), is an important quantity for signal analysis and the design of RF transmitters and receivers.  Modern digital instrumentation, like vector signal analyzers, collect and process discrete-time samples of the complex baseband signal.  Also, computer-based simulations of continuous-time processes must be carried out with discrete-time sequences.   Therefore, to design such measurements and simulations, accurate models for the distribution of the PAPR for discrete-time complex baseband processes are required.  In particular, band-pass thermal noise and modern communication signals, like orthogonal frequency division multiplexing (OFDM), are often modeled as complex Gaussian processes.  To address this problem, we present a new analytical model for the PAPR distribution of a stationary, discrete-time, complex Gaussian process.  We demonstrate that the new discrete-time model subsumes prior models based on asymptotic extreme value theory as special cases while providing a larger domain of efficacy.  Specifically, using Monte Carlo simulations of filtered complex white Gaussian noise, we show that the new model is more effective for practical sample sizes and dual-band spectra than previous approaches.   
\end{abstract}


\section{Introduction}
\label{sec:introduction}

It is common in radio frequency (RF) measurements to record the peak instantaneous power observed over a specified time interval.  For example, assessments of RF spectrum utilization often perform peak detection to capture transient or short-duration attributes and to distinguish emission types \cite{Cotton2014,Kuester2022,Sanders2026}.  Peak power measurements are also used in the characterization of transmitters for compliance testing \cite{ANSI_c63_26}.  These applications involving band-limited, modulated RF signals are typically implemented using modern digital instrumentation, such as a vector signal analyzer, that collects sampled, discrete-time measurements of the complex baseband signal, also known as the in-phase and quadrature (I/Q) representation \cite{Proakis2002, Lapidoth2017}. 

In addition, modeling and reducing the peak-to-average power ratio (PAPR) of continuous-time, analog signals is important in communication system design to avoid impacts of power amplifier nonlinearities on RF transmission \cite{Rahmatallah2013, Nikandish2020}.  Consequently, a large body of literature has investigated models for continuous-time PAPR and PAPR reduction schemes for communication signals, particularly for  orthogonal frequency division multiplexing (OFDM) signals \cite{Rahmatallah2013}.  Assuming that the baseband OFDM signal can be modeled as a complex Gaussian process \cite{Wei2010}, several works have applied extreme value theory to derive expressions for the PAPR distribution of continuous-time OFDM signals and verified their results with discrete-time simulations \cite{Ochiai2001,Wei2002,Wei2003_thesis,Jiang2008}.  

Because complex Gaussian processes originate from (i) band-pass thermal noise that is present in all baseband RF measurements \cite{Vasilescu2006}, and (ii) communication signals implementing OFDM \cite{Wei2010}, accurate mathematical models for the peak power distribution of discrete-time complex Gaussian processes are valuable for the design of measurements and simulations. Established extreme value theory relevant to this problem includes the general theory for stationary, dependent random sequences \cite{Leadbetter1983,Leadbetter1988,Smith2025} and results for the envelope of a continuous-time Gaussian process \cite{Cramer1967, Vanmarcke2010, Leadbetter1983,Leadbetter1988}.  

\IEEEpubidadjcol 

The general theory for extreme values of stationary discrete-time random sequences focuses on technical conditions under which samples are asymptotically independent and high-level exceedances (level crossings) do not cluster.  When such conditions are satisfied, the asymptotic extreme value distribution is the same as that for an independent and identically distributed (i.i.d.) sequence with the same marginal distribution \cite{Leadbetter1983,Leadbetter1988,Smith2025}.  While mathematically correct, the general asymptotic theory for random sequences unfortunately does not explain how, for practical, finite sample sizes, the extreme value distribution depends on spectral characteristics of the random process, which is required for practical engineering applications.  For example, in prior work \cite[Sec.~IV-C]{Wunderlich2025}, we assessed the mean PAPR of low-pass filtered complex white Gaussian noise (CWGN).  For sequence lengths less than $10^5$, we found that the mean PAPR is well approximated by the i.i.d. model when the low-pass filter cutoff frequency is $\geq 0.25$ cycles per sample, but poorly approximated by the i.i.d. model when the cutoff is $\leq 0.1$ cycles per sample.

Alternatively, results for the envelope of a continuous-time Gaussian process \cite{Cramer1967,Vanmarcke2010,Leadbetter1983,Leadbetter1988} can be utilized to obtain asymptotically valid expressions that can be discretized.  This approach has the advantage that it yields theoretical expressions that depend on the spectral attributes of the underlying random process.  

As supported by the aforementioned literature on continuous-time OFDM modeling \cite{Tellambura2001, Ochiai2001, Wei2002, Wei2003_thesis, Jiang2008}, PAPR distributions derived from continuous-time theory are known to agree with those for corresponding sampled low-pass processes when the oversampling ratio (OSR) is sufficiently large.  Specifically, prior studies used an OSR as low as four \cite{Tellambura2001} and and high as thirty-two \cite{Wei2003_thesis}.  This wide range illustrates that the degree of oversampling necessary to accurately recover continuous-time PAPR distributions in discrete-time simulation is not well understood. Investigations studying recovery of peak values for sampled deterministic signals include \cite{Wunder2003}, which gave tight bounds relating peak values of continuous-time and oversampled discrete-time signals, and \cite{Boche2020}, which established the necessity of oversampling.  

Thus, there is evidence of at least two fundamental regimes for the peak power distribution of a sampled continuous-time complex Gaussian process: one for smaller OSRs, asymptotically matching the distribution for an i.i.d. sequence, and one for larger OSRs, that is well approximated by the distribution for the continuous-time process.  Indeed, related theoretical work on sampled $\chi$-processes \cite{Ling2016} has shown that under certain conditions, the extreme value distribution is asymptotically equivalent to that for an i.i.d. sequence when the sampling grid is ``sparse'' and asymptotically equivalent to the distribution implied by continuous-time theory when the sampling grid is ``dense,'' with a transitional regime between the two.  However, because the definitions of ``sparse'' and ``dense'' sampling grids in \cite{Ling2016} are inherently asymptotic, it is not clear how they could be applied in practice for a specific sampling period and observation interval.    

\begin{figure}[tb]
    \centering
    \subfloat[Filter frequency response.]{%
        \includegraphics[width = 0.48\columnwidth]{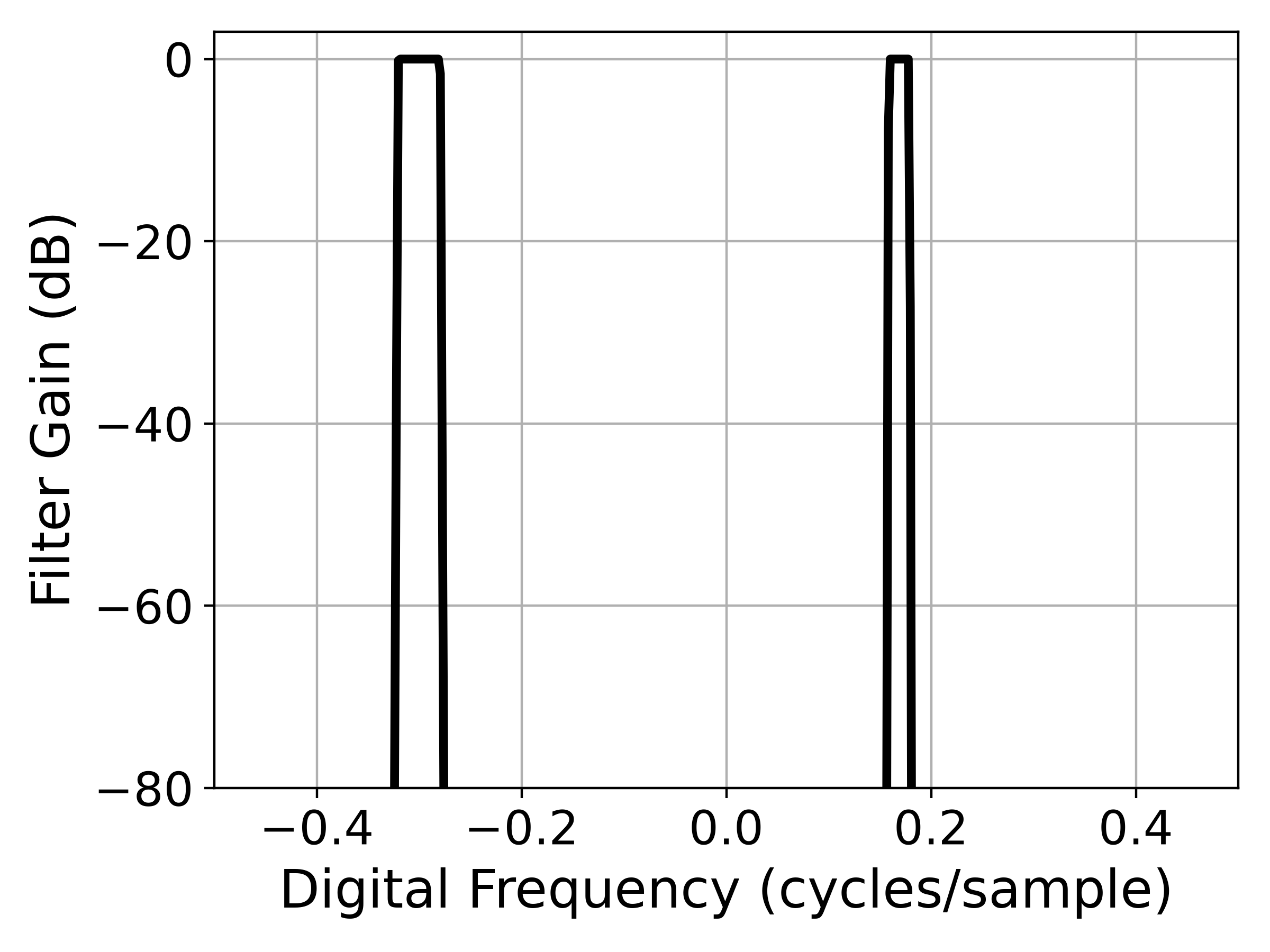}
    }
    \hfill
    \subfloat[Probability density functions.]{%
        \includegraphics[width = 0.48\columnwidth]{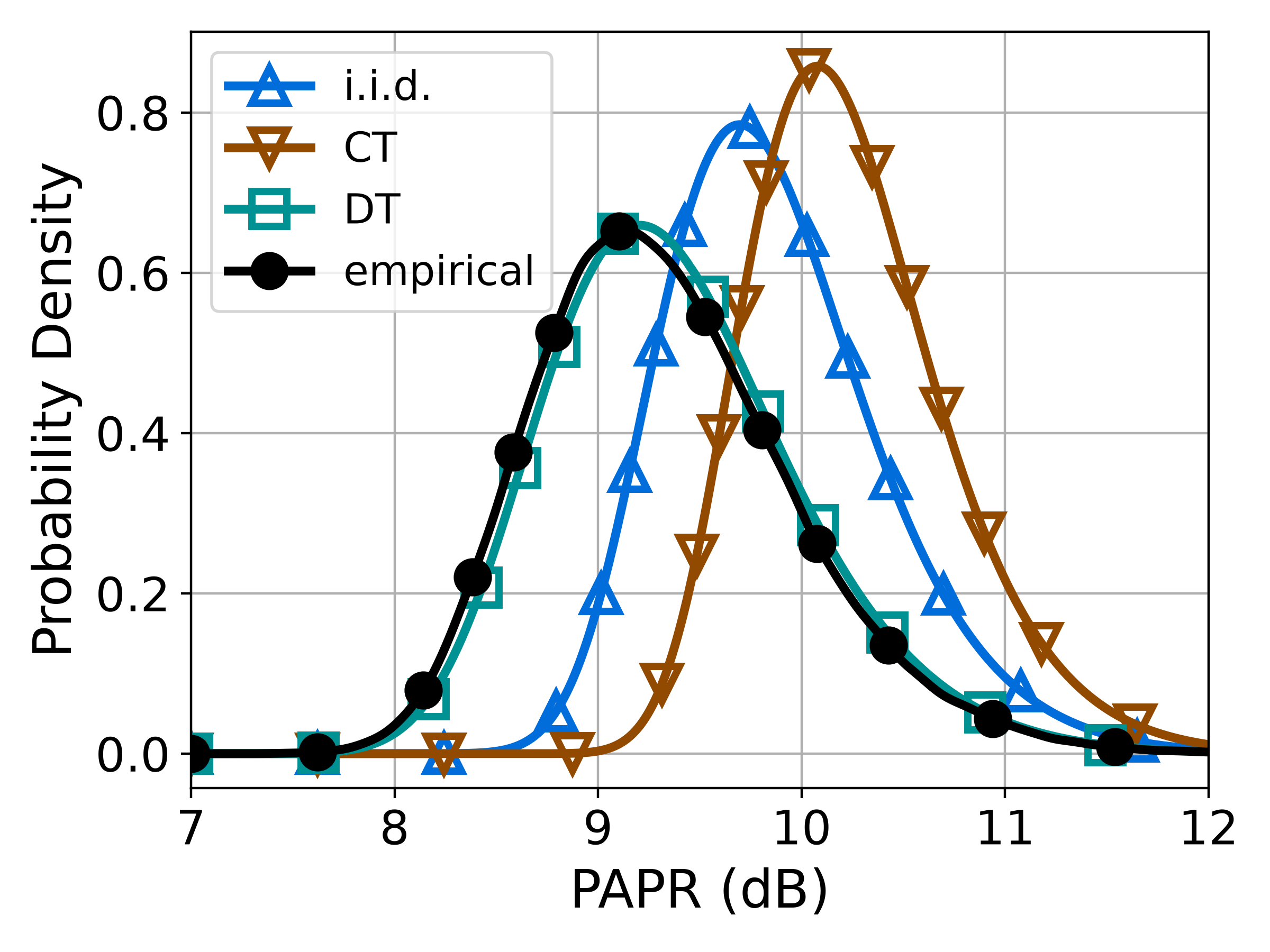}
    }
    \caption{PAPR distribution for a filtered CWGN sequence of length $10^4$ with asymmetric, dual-band spectrum.  The empirical distribution was estimated from $10^5$ independent Monte Carlo trials.  The labels ``i.i.d.'', ``CT'', and ``DT'' identify the i.i.d. model, continuous-time model, and our new discrete-time model, respectively. Corresponds to $\mathrm{OSR}_N = 2$, $\mathrm{OSR}_L = 16$ in Fig.~\ref{fig:asymmetric_2comp_BP}.}
    \label{fig:teaser_plot}
\end{figure}

In this paper, we present a new analytical, discrete-time model for the PAPR distribution of a sampled complex baseband Gaussian process. Our approach is based on a first-order, two-state Markov model for level exceedances \cite{Vanmarcke1975,Morgan2007} and established results on the bivariate Rayleigh distribution \cite{Schwartz1966,Simon2005}.  We show that the new discrete-time model subsumes the i.i.d. and discretized continuous-time models as special cases and is effective over a wider range of scenarios.  The efficacy of the new model is illustrated in Fig.~\ref{fig:teaser_plot}, which presents a challenging dual-band scenario with filtered CWGN.  

The paper is organized as follows.  We begin in Sec.~\ref{sec:foundations} by reviewing necessary theory for complex Gaussian baseband random processes in both continuous-time and discrete-time, including distribution theory relevant to the instantaneous power.  Next, in Sec.~\ref{sec:current_theory}, we summarize established results from extreme value theory, including the PAPR distribution for i.i.d. samples and results for the envelope of a continuous-time Gaussian process.  Further details on the continuous-time theory and connections to the literature on OFDM PAPR modeling are given the Appendix~\ref{appendix:CT_PAPR_theory}.  Our main result is presented in Sec.~\ref{sec:practical_DT_PAPR_model}, with additional theoretical derivations provided in Appendices~\ref{appendix:bivariate_exponential} and \ref{appendix:equiv_CTDT_DT}.  Evaluation methods are detailed in Sec.~\ref{sec:eval_methods} and simulation results are presented in Sec.~\ref{sec:results}.  Last, Sec.~\ref{sec:conclusions} contains the discussion and conclusions.

\section{Foundations}
\label{sec:foundations}

In this section, we lay the groundwork for later developments by introducing the problem setting and notation, and reviewing important properties.  We start in the continuous-time setting and then cover adjustments to notation for the corresponding discrete-time domain.   

Background on random processes and their applications to communication engineering can be found in many textbooks, such as \cite{WozencraftJacobs1965, Schwartz1966, Proakis2002, Lapidoth2017}.  We assume throughout that appropriate technical mathematical requirements are satisfied so that all quantities are well-defined and mathematical steps are valid.

\subsection{Complex Gaussian Baseband Processes}
\label{sec:complex_baseband_processes}

Let $X(t)$ be a continuous-time, real, passband Gaussian random process with central frequency, $f_0$, and half-bandwidth, $W$, that is band-limited to $f_0 \pm W$~Hz, where $W < f_0$.  Specifically, the passband process is assumed to be wide-sense stationary (WSS) and zero-mean, taking the form 
\begin{equation}
    X(t) = I(t)\cos(2\pi f_0 t) - Q(t)\sin(2\pi f_0 t), 
\end{equation}
where $I(t)$ and $Q(t)$ are the real-valued in-phase and quadrature components of the complex baseband process $Z(t) = I(t) + jQ(t)$ with $j^2 = -1$ \cite{Proakis2002, Lapidoth2017}.  The zero-mean baseband process, $Z(t)$, can be interpreted as the complex envelope of the passband process, $X(t)$.  Furthermore, when $W < f_0$, the ``natural'' complex envelope, $Z(t)$, is equivalent to the ``analytic'' complex envelope defined via the analytic signal representation of the passband process \cite{Rice1982, Lapidoth2017}.  

Under the above assumptions, the baseband process $Z(t)$ can be shown to be a zero-mean, WSS, proper complex Gaussian random process\footnote{Furthermore, a WSS proper complex Gaussian process is circularly symmetric and strictly stationary \cite[Appendix~D]{Lapidoth2017}.} \cite[Thm D.6]{Lapidoth2017} \cite[Thm.~3]{Neeser1993}.  Letting $E[\cdot]$ denote mathematical expectation and using a star to represent complex conjugation, define the autocorrelation function (ACF) as $r_{ZZ}(\tau) = E[Z(t+\tau)Z^*(t)]$ and the pseudo-autocorrelation function (pseudo-ACF) as $\widetilde{r}_{ZZ}(\tau) = E[Z(t+\tau)Z(t)]$.  Above, the terminology ``proper'' indicates that the pseudo-ACF vanishes\footnote{More precisely, the pseudo-autocovariance function vanishes for a proper complex random process.}, i.e., $\widetilde{r}_{ZZ}(\tau) = 0$ \cite{Neeser1993}.

Let the ACFs for the real-valued in-phase and quadrature baseband components be $r_{II}(\tau) = E[I(t+\tau)I(t)]$ and $r_{QQ}(\tau) = E[Q(t+\tau)Q(t)]$, and the cross-correlation functions (CCFs) be $r_{IQ}(\tau) = E[I(t+\tau)Q(t)]$ and $r_{QI}(\tau) = E[Q(t+\tau)I(t)]$. It is straightforward to see that the pseudo-ACF, $\widetilde{r}_{ZZ}(\tau) = E[Z(t+\tau)Z(t)]$, can be expressed as
\begin{equation}
    \widetilde{r}_{ZZ}(\tau) = [r_{II}(\tau) - r_{QQ}(\tau)] + j[r_{QI}(\tau) + r_{IQ}(\tau)].
    \label{eq:Rzz_tilde_decomp}
\end{equation}

Because the complex baseband process is proper, the pseudo-ACF vanishes, and (\ref{eq:Rzz_tilde_decomp}) yields \cite[Thm.~3]{Neeser1993} \cite[p.~498]{WozencraftJacobs1965}  
\begin{equation}
    r_{II}(\tau) = r_{QQ}(\tau) \quad \text{and} \quad r_{QI}(\tau) = -r_{IQ}(\tau).
    \label{eq:cc_proper}
\end{equation}
The first equation implies that $I(t)$ and $Q(t)$ are equivalent in distribution, since they are both zero-mean Gaussian processes.  Further, since $r_{QI}(\tau) = r_{IQ}(-\tau)$, the second equation implies that $r_{IQ}(-\tau) = -r_{IQ}(\tau)$ and $r_{QI}(\tau) = -r_{QI}(-\tau)$, so the CCFs are both odd functions.  Specifically, $r_{IQ}(0) = r_{QI}(0) = 0$, so $I(t)$ and $Q(t)$ are uncorrelated, and hence independent due to being Gaussian, at each time index, $t$.  Note however, that because the CCFs are generally nonzero for nonzero lags, $I(t)$ and $Q(t)$ are not generally independent random processes. 

The complex baseband process can also be characterized in terms of its power spectral density (PSD), $S_{ZZ}(f) =  \int_{-\infty}^{\infty} r_{ZZ}(\tau)e^{-j2\pi f \tau}d\tau$.  In particular, the first three spectral moments,
\begin{equation}
    \lambda_k = \int_{-\infty}^{\infty} (2\pi f)^k S_{ZZ}(f)\,df, \qquad k=0,1,2
\end{equation}
will be used later.  Throughout this work, we assume that the baseband process has finite second spectral moment, i.e., $\lambda_2 < \infty$, which implies that $Z(t)$ is mean square differentiable and that $\lambda_1$ and $\lambda_0$ are finite \cite{Cramer1967}.  Since $r_{ZZ}(\tau)$ and $S_{ZZ}(f)$ are a Fourier transform pair, it follows from Fourier transform properties that the spectral moments can be evaluated in terms of the ACF and its derivatives at $\tau=0$.  Namely, $\lambda_0 = r_{ZZ}(0)$, $\lambda_1 = -j r^{\prime}_{ZZ}(0)$, and $\lambda_2 = - r^{\prime \prime}_{ZZ}(0)$.  Note that $\lambda_0 = \text{Var}[Z(t)]$, and $\lambda_1 = 0$ if and only if $S_{ZZ}(f)$ is an even function of frequency.

Expanding $r_{ZZ}(\tau)$ in terms of the I/Q component correlation functions and applying (\ref{eq:cc_proper}) yields 
\begin{equation}
    r_{ZZ}(\tau) = 2r_{II}(\tau) - j2r_{IQ}(\tau).
    \label{eq:z_acf}
\end{equation}
By elementary Fourier transform properties, the real and imaginary parts of $r_{ZZ}(t)$, are the inverse Fourier transforms of the even and odd parts of $S_{ZZ}(f)$. Thus, $S_{ZZ}(f)$ is a strictly even function of frequency ($\lambda_1=0$), if and only if $r_{IQ}(\tau) = 0$, which is equivalent to $I(t)$ and $Q(t)$ being independent random processes.  

The above definitions and properties carry over to the discrete-time setting, where we assume that the continuous-time baseband process, $Z(t)$, is sampled uniformly at a rate, $f_s$, with $f_s > 2W$ to avoid aliasing. For integer indices $i$ and $k$, denote the sampled, discrete-time, WSS complex baseband process as $Z_i = I_i + jQ_i$, its ACF as $r_{ZZ}[k] =E[Z_{i+k}Z^*_i]$, and its pseudo-ACF as $\widetilde{r}_{ZZ}[k] =E[Z_{i+k}Z_i]$.  Similar to the continuous-time case, $Z_i$ is a zero-mean, proper complex Gaussian random process, which implies that $\widetilde{r}_{ZZ}[k] = 0$.  

The PSD for the discrete-time process is denoted $S_{ZZ}^d(f_d) = \sum_{k=-\infty}^{\infty} r_{ZZ}[k]e^{-j2\pi f_d k}$ where $f_d = f/f_s \in [-1/2, 1/2)$ is normalized digital frequency expressed in cycles per sample, and the superscript ``$d$'' indicates that the PSD is for the discrete-time process.  Since there is no aliasing, standard sampling theory implies that the discrete-time and continuous-time PSDs are related by \cite[Eq. (10.50)]{Oppenheim1999}
\begin{equation}
    S_{ZZ}^d(f_d) = f_sS_{ZZ}(f_d f_s) \qquad \text{for } -1/2 \leq f_d < 1/2.
    \label{eq:DTCT_PSD_relation}
\end{equation}
Defining spectral moments for the discrete-time process as
\begin{equation}
    \lambda_k^d = \int_{-1/2}^{1/2} (2\pi f_d)^k S_{ZZ}^d(f_d)\,df_d \qquad \text{for } k=0,1,2
\end{equation}
it follows from (\ref{eq:DTCT_PSD_relation}) that the discrete-time and continuous-time spectral moments are related by
\begin{equation}
    \lambda_k^d = \lambda_k/f_s^k.
    \label{eq:lambda_conversion}
\end{equation}
In particular, since $\lambda_0 = \lambda_0^d$, we will use $\lambda_0$ to denote the zeroth spectral moment in both discrete and continuous time.  

\subsection{Instantaneous Power, Peak-to-Average-Power Ratio, and Related Quantities}
\label{sec:foundations_PAPR}

Let $P(t) = |Z(t)|^2= I^2(t) + Q^2(t)$ be the instantaneous power, or squared amplitude, of the complex baseband Gaussian process.  Observe that the random processes $P(t)$ is invariant to frequency translations of $Z(t)$.  Namely, let $\widehat{Z}(t) = Z(t) \exp(j 2 \pi f_0 t)$.  Since $P(t) = |Z(t)|^2 = |\widehat{Z}(t)|^2$ for every realization (sample path) of $Z(t)$, each realization of $P(t)$ is unchanged, and the random process $P(t)$ is invariant to arbitrary frequency shifts of the complex baseband process, $Z(t)$.  This fundamental fact implies that the extreme value distribution of $P(t)$ will likewise not change.  

From the prior subsection, at each time $t$, $I(t)$ and $Q(t)$ are i.i.d. normal random variables with zero mean and variance $\lambda_0/2$, denoted $I(t) \sim \mathcal{N}(0,\lambda_0/2)$ and $Q(t) \sim \mathcal{N}(0,\lambda_0/2)$.  Also, $Z(t)$ is a proper complex normal random variable with zero mean and variance $\lambda_0$, denoted $Z(t) \sim \mathcal{CN}(0,\lambda_0)$.    

The marginal distribution of $P(t)$ at each time $t$ can be shown to be an exponential distribution, with cumulative distribution function (CDF) $F_P(x) = 1-\exp(-x/\lambda_0)$ and rate parameter, $1/\lambda_0$, which we denote as $P(t) \sim \text{Exp}(1/\lambda_0)$.  Equivalently, $2P(t)/\lambda_0$ has a $\chi^2$ marginal distribution with two degrees of freedom, denoted $\chi_2^2$ \cite{Johnson1994}.  As discussed in the prior subsection, $I(t)$ and $Q(t)$ are independent random processes, and hence, $2P(t)/\lambda_0$ is a $\chi^2$ random process, if and only if $\lambda_1=0$.  This distinction is necessary to understand prior literature on $\chi^2$ and $\chi$ processes, discussed below.      

For a finite time interval $[0,T]$, the PAPR is defined as
\begin{equation}
    \text{PAPR} = \frac{\max_{t \in [0,T]} P(t)}{E[P(t)]} = \frac{M_T}{\lambda_0},
\end{equation}
where $M_T = \max_{[0,T]} P(t)$.  Note that PAPR is invariant to multiplicative scaling of $Z(t)$.  Therefore, without loss of generality, unless stated otherwise, we assume that $Z(t)$ is scaled so that $\lambda_0 = \text{Var}[Z(t)] = 1$.  In this case, the marginal distributions are standardized, i.e., $Z(t) \sim \mathcal{CN}(0,1)$ and $P(t) \sim \mathrm{Exp}(1)$, and PAPR takes the simpler form $\text{PAPR} = \max_{t \in [0,T]} P(t)$.  

In some contexts, because the median is a more robust measure central tendency than the mean, it is preferable to consider the peak-to-median power ratio (PMPR) \cite{Sanders2026b}. For the standard exponential distribution, the mean is one and median is $\log(2)$, where we write $\log(\cdot)$ for natural logarithm unless specified otherwise.  Therefore, all results presented here for PAPR apply to PMPR with the transformation $\text{PAPR} = \text{PMPR}(\log 2)$.

A quantity closely related to PAPR is the crest factor (CF), also known as peak factor, which is defined as the peak-to-average ratio of the amplitude, or real-valued envelope, $R(t) = \sqrt{I^2(t) + Q^2(t)}$, and hence, $\text{CF} = \sqrt{\text{PAPR}}$.  When $P(t) \sim \text{Exp}(1/\lambda_0)$, the marginal distribution of $R(t)$ is a Rayleigh distribution with scale parameter $\sigma = \sqrt{\lambda_0/2}$, denoted $R(t) \sim \text{Rayleigh}(\sqrt{\lambda_0/2})$, and CDF $F_R(x) = 1 - \exp(-x^2/\lambda_0)$.  

In discrete-time, the power at integer index $i$ is $P_i = |Z_i|^2$ and PAPR is defined for a sequence of size $n$ as
\begin{equation}
    \text{PAPR} = \frac{\max_{1 \leq i \leq n} P_i}{E[P_i]} = \frac{M_n}{\lambda_0}
\end{equation}
where $M_n = \max_{1 \leq i \leq n} P_i$. Similar to the continuous-time case, because the random process $P_i$ is invariant to frequency translation of $Z_i$, the distribution of PAPR is likewise invariant.\footnote{Because the discrete-time frequency domain is topologically equivalent to a 1-D torus, or circle, this property applies to any frequency translation of the PSD, $S_{ZZ}^d(f_d)$, modulo $1$.} Also, similar to the continuous-time setting, the marginal distribution of $P_i$ is exponential with rate parameter $1/\lambda_0$, denoted $P_i \sim \text{Exp}(1/\lambda_0)$, and the discrete-time amplitude, $R_i$, has a marginal Rayleigh distribution, $R_i \sim \text{Rayleigh}(\sqrt{\lambda_0/2})$.

\section{Current Theory}
\label{sec:current_theory}
Here, we review results from the literature on extreme value theory that can be applied to model the PAPR distribution of a WSS complex baseband Gaussian process in discrete-time.  

As highlighted in the prior section, the condition $\lambda_1=0$ is necessary and sufficient for the baseband components, $I(t)$ and $Q(t)$, to be independent random processes.  A large portion of the pertinent extreme value literature, particularly on $\chi^2$ and $\chi$ processes \cite{Leadbetter1988,Sharpe1978,Lindgren1980,Ling2016} and OFDM \cite{Ochiai2001,Wei2002,Jiang2008}, assumes that the baseband component processes are independent.  For this reason, we will be careful to specify when prior literature addresses only the $\lambda_1=0$ case.            

The general theory for extreme values of stationary discrete-time random sequences focuses on technical conditions under which samples are asymptotically independent and high-level exceedances (level crossings) do not cluster.  When such conditions are satisfied, the asymptotic extreme value distribution is the same as for an i.i.d. sequence with the same marginal distribution \cite{Leadbetter1983,Leadbetter1988,Smith2025}.  

Therefore, we begin by reviewing the PAPR distribution predicted by classical extreme value theory for an i.i.d. sequence.  Specifically, suppose the power sequence, $P_1, P_2, \ldots, P_n$, is i.i.d. with an standard exponential distribution, i.e., $P_i \sim \text{Exp}(1)$.  Then the exact PAPR CDF is \cite[Ch.~2]{Castillo1988}  
\begin{equation}
F_{\text{PAPR}}^{\text{iid}}(x) = [F_P(x)]^n = [1-\exp(-x)]^n,
\label{eq:PAPR_iid}
\end{equation}
where the superscript ``iid'' indicates that the CDF is for an i.i.d. sequence.  From classical extreme value theory, it follows that (\ref{eq:PAPR_iid}) is asymptotically equivalent to 
\begin{equation}
    F_{\text{PAPR}}^{\text{iid}}(x) \approx \exp[-\exp(-(x - \log n))]
    \label{eq:PAPR_iid_gumbel}
\end{equation}
as $n \rightarrow \infty$ \cite[Example~3.5]{Castillo1988}.  Eq. (\ref{eq:PAPR_iid_gumbel}) has the form of a Gumbel distribution with CDF $F(x; a_n, b_n) = \exp[-\exp[-a_n(x-b_n)]]$, where $a_n=1$ is a scale parameter and $b_n = \log n$ is a location parameter \cite{Castillo1988}.  Additional results for the i.i.d. case are given in \cite{Wunderlich2025}.   

To our knowledge, theoretical results on sufficient conditions for the extreme value distribution of the sequence, $P_i$, to be asymptotically equivalent to that for an i.i.d. sequence are not available in the general case when $\lambda_1 \neq 0$. However, in the context of sampled continuous-time $\chi$-processes, where $\lambda_1 = 0$, Ling and Tan \cite{Ling2016} have shown that the so-called Berman condition,  which in our notation is $r_{II}(\tau) \log \tau \rightarrow 0$ as $\tau \rightarrow \infty$,\footnote{Any continuous-time, WSS random process with ACF that decays as $\mathcal{O}(1/|\tau|)$ satisfies the Berman condition, where $\mathcal{O}(\cdot)$ denotes Big-O notation.  Sufficient conditions for this property are that the PSD is integrable and has bounded variation \cite[Thm.~1]{Ghodadra2015}. Most processes of practical interest satisfy these requirements.} is sufficient for the extreme value distribution to be asymptotically equivalent to that for an i.i.d. sequence when the sampling grid is ``sparse.'' Furthermore, Ling and Tan show that for a ``dense'' sampling grid, the asymptotic distribution implied by continuous-time theory applies.\footnote{See \cite[Thm~2.1]{Ling2016} with $\alpha=2, m=0, r=0$.}  

We do not review Ling and Tan's results in detail here, but note that the definitions of a ``sparse'' or ``dense'' sampling grid, which are drawn from prior work by Piterbarg \cite{Piterbarg2004} on Gaussian processes, are inherently asymptotic statements relating the sampling period and the observation interval.  As such, it is not evident how the notion of a ``sparse'' or ``dense'' grid could be inferred for a specific sampling rate and observation interval in practice.  Nonetheless, the theory of Ling and Tan \cite{Ling2016} clarifies that there are three fundamental asymptotic regimes for the extreme value distribution in discrete-time: one corresponding to the i.i.d. case (``sparse'' grid), one corresponding to the continuous-time case (``dense'' grid), and a transitional regime called a ``Pickands'' grid.  
     
Continuous-time extreme value theory for PAPR in our problem setting and connections to the communications engineering literature are reviewed in Appendix~\ref{appendix:CT_PAPR_theory}.  In particular, under the generalized Berman condition $r_{ZZ}(\tau) \log(\tau) \rightarrow 0$ as $\tau \rightarrow \infty$, the CDF for PAPR, $F_{\text{PAPR}}(x) = P(\text{PAPR} \leq x)$, is approximated by the asymptotically equivalent expressions (\ref{eq:PAPR_CDF_CT_1}) and (\ref{eq:PAPR_CDF_CT_2}), where $T_0$ is a reference time base.  We proceed with a discretization of the second expression, (\ref{eq:PAPR_CDF_CT_2}), since it can be expressed in the form of a Gumbel distribution, which has well-known properties.  

Suppose that the continuous-time complex baseband process, $Z(t)$, is sampled uniformly over the time interval $[0,T)$ at $f_s = 1/T_s$ samples per second to obtain the discrete-time sequence, $Z_i$.  Namely, samples are taken at $t_i = (i-1)T_s$ for $i=1, \ldots, n$, where $n = T/T_s$ is a positive integer.  Letting the reference time base be $T_0 = T_s$, it follows from (\ref{eq:PAPR_CDF_CT_2}) that the PAPR CDF can be modeled as
\begin{equation}
    F_{\text{PAPR}}^{\text{CT}}(x) =  \exp\left[-ne^{-x}\sqrt{\frac{\lambda^d_2 - (\lambda^d_1)^2}{\pi}\log n}~\right],
    \label{eq:PAPR_CDF_CTDT}
\end{equation}
where we used (\ref{eq:lambda_conversion}) to relate the spectral moments in continuous-time and discrete-time.  The superscript ``CT'' indicates that the CDF is derived from continuous-time theory.  

Eq. (\ref{eq:PAPR_CDF_CTDT}) depends on the PSD, $S_{ZZ}^d(f_d)$, through the spectral dispersion, $\lambda^d_2 - (\lambda^d_1)^2$.  It is straightforward to see that the spectral dispersion, and hence (\ref{eq:PAPR_CDF_CTDT}), does not change when $S_{ZZ}^d(f_d)$ is shifted in frequency, provided the support of the shifted PSD is not split across the Nyquist boundary, $f_d = 1/2$.

Observe that (\ref{eq:PAPR_CDF_CTDT}) can be rewritten as a Gumbel distribution, $F_{\text{PAPR}}^{\text{CT}}(x) \approx \exp[-\exp[-a_n(x-b_n)]]$, with scale parameter, $a_n=1$, and location parameter, 
\begin{equation}
    b_n = \log n + \frac{1}{2}\log\log n  +\log\left(\sqrt{\frac{\lambda^d_2 - (\lambda^d_1)^2}{\pi}}\right). 
\end{equation}
Thus, comparing the Gumbel distribution parameters of (\ref{eq:PAPR_CDF_CTDT}) with those of (\ref{eq:PAPR_iid_gumbel}) for the i.i.d. case, we see that the scale parameter is the same, but the location parameter includes additional terms reflecting sample dependence.

\section{A Practical Analytical Model for the Discrete-Time PAPR Distribution}
\label{sec:practical_DT_PAPR_model}

We will see below in Sec.~\ref{sec:results} that the ranges of applicability for (\ref{eq:PAPR_iid_gumbel}) and (\ref{eq:PAPR_CDF_CTDT}) are complementary, but limited.  For this reason, we introduce a new, fully discrete-time, non-asymptotic model that directly models sample dependence.  Specifically, we model the level crossings (or exceedances) of the power sequence, $P_i$, as a first-order, two-state Markov chain.  This assumption is motivated by prior work of Vanmarcke \cite{Vanmarcke1975} on continuous-time extremes for Gaussian processes and also by a study of Morgan \cite{Morgan2007} concerning the level crossing rate and sojourn probability for low-pass Gaussian processes in both discrete-time and continuous-time.  

\subsection{Discrete-time Markov Model}
\label{sec:markov_model}

For time index, $i$, and level, $x$, define the binary exceedance sequence, $Y_i = \mathcal{I}(P_i > x)$, where $\mathcal{I}(S)$ is the indicator function that equals one if the proposition $S$ is true and equals zero otherwise.  So $Y_i = 0$ in the event of non-exceedance, $P_i \leq x$, and $Y_i=1$ in the event of exceedance, $P_i > x$.  

Suppose that $Y_i$ can be modeled as a stationary, first-order, two-state Markov chain.  In this case, the probability of the next state depends only on the current state, i.e., $P(Y_{i+1}=y_{i+1} \mid Y_i=y_i, Y_{i-1}=y_{i-1},\ldots) = P(Y_{i+1}=y_{i+1} \mid Y_i=y_i)$ where $y_i \in \{0,1\}$.  We denote the transition probability of moving from state $\alpha \in \{0, 1\}$ to state $\beta \in \{0, 1\}$ as $p_{\alpha\beta}(x) = P(Y_{i+1}=\beta \mid Y_i=\alpha)$.   For example, the probability of a level upcrossing is $p_{01}(x) = P(Y_{i+1} = 1 \mid Y_i = 0) = P(P_{i+1} > x \mid P_i \leq x)$.  By the law of total probability, it follows that $p_{00}(x) = 1 - p_{01}(x)$ and $p_{11}(x) = 1-p_{10}(x)$.  Also, denote the stationary probabilities for each state as $\pi_0(x) = P(Y_i = 0) = P(P_i \leq x)$ and $\pi_1(x) = P(Y_i=1) = P(P_i > x)$.  

Recalling the notation $M_n = \max_{1 \leq i \leq n} P_i$ from Sec.~\ref{sec:foundations_PAPR}, the above Markov assumptions imply that 
\begin{align}
    P(M_n \leq x) &= P(P_1 \leq x, P_2 \leq x, \ldots, P_n \leq x) \\
    &= P(P_1 \leq x) P(P_2 \leq x \mid P_1 \leq x) \nonumber \ldots \\ &\quad\times P(P_n \leq x \mid P_{n-1} \leq x) \\ 
    &= \pi_0(x) p_{00}^{n-1}(x).
\end{align}
Further, scaling $Z_i$ so that $\lambda_0 = \text{Var}[Z_i] = 1$, it follows without loss of generality, that the CDF for PAPR is 
\begin{equation}
    F_{\text{PAPR}}^{\text{DT}}(x) = \pi_0(x) p_{00}^{n-1}(x),
\label{eq:CDF_DT}
\end{equation}
where the superscript ``DT'' indicates that the CDF corresponds to the discrete-time, first-order, two-state Markov model for level exceedances.  From Sec.~\ref{sec:foundations_PAPR}, we have $\pi_0(x) = P(P_i \leq x) = 1 - \exp(-x)$.  We present an analytical expression for $p_{00}(x)$ below.

\subsection{Main Result}

To derive formulas for the state transition probabilities, and specifically, $p_{00}(x)$, we require the joint CDF for $P_i$ and $P_{i+1}$, which each have a standard exponential marginal distribution.  Fortunately, the joint CDF can be obtained from established results for the bivariate Rayleigh distribution \cite[Appendix~A]{Schwartz1966} \cite[Sec.~6.1]{Simon2005}.  In particular, as outlined in Appendix~\ref{appendix:bivariate_exponential}, defining $\kappa = |r_{ZZ}[1]|$, we find that for $\lambda_0=1$ and  $0 \leq \kappa < 1$,
\begin{equation}
        P(P_i \leq x, P_{i+1} \leq x) = 1-e^{-x} -  e^{-x}\Psi(x,\kappa),
       \label{eq:joint_P_CDF_xx}
\end{equation}
where $\Psi(x,\kappa)$ takes the alternative forms
\begin{align}
\Psi(x,\kappa) &= Q_1(c,\kappa c) - Q_1(\kappa c, c) \label{eq:Psi_Q} \\
&= 1 - \frac{1}{2\pi} \int_{-\pi}^{\pi}\exp\left[ \frac{-x(1-\kappa^2)}{1 + 2\kappa \sin \theta +\kappa^2}\right]d\theta. \label{eq:Psi_integral}
\end{align}
Above, $c = \sqrt{2x/(1-\kappa^2)}$, and $Q_1$ is the first-order Marcum-Q function \cite{Schwartz1966, Simon2005}.  Because (\ref{eq:Psi_Q}) requires evaluation of the Marcum-Q function, we prefer to use (\ref{eq:Psi_integral}), as explained below.  

From (\ref{eq:joint_P_CDF_xx}), the probability, $p_{00}(x)$ is found to be
\begin{align}
    p_{00}(x) &= P(P_{i+1} \leq x \mid P_i \leq x) \\
    &= \frac{P(P_i \leq x, P_{i+1} \leq x)}{P(P_i \leq x)} \\
    &= 1 - \left(\frac{e^{-x}}{1-e^{-x}}\right) \Psi(x,\kappa)  \label{eq:p_00}
\end{align}
Finally, (\ref{eq:CDF_DT}) and (\ref{eq:p_00}) yield the PAPR CDF
\begin{equation}
    F_{\text{PAPR}}^{\text{DT}}(x)  = (1 - e^{-x})\left[ 1 - \left(\frac{e^{-x}}{1-e^{-x}}\right) \Psi(x,\kappa)\right]^{n-1}.
    \label{eq:PAPR_CDF_DT_full}
\end{equation}

The PAPR CDF given by (\ref{eq:PAPR_CDF_DT_full}) is our main result.  Below, in Sec.~\ref{sec:results}, we will see that it accurately models the PAPR distribution over a broader range of conditions than (\ref{eq:PAPR_iid}) and (\ref{eq:PAPR_CDF_CTDT}).  Indeed, it can be seen that (\ref{eq:PAPR_CDF_DT_full}) includes (\ref{eq:PAPR_iid}) and (\ref{eq:PAPR_CDF_CTDT}) as special cases.  Namely, observe that when $Z_i$ and $Z_{i+1}$ are i.i.d., $\kappa = 0$ and $\Psi(x,\kappa) = 1-\exp(-x)$. So (\ref{eq:PAPR_CDF_DT_full}) reduces to (\ref{eq:PAPR_iid}) in the i.i.d. case.  Additionally, in Appendix~\ref{appendix:equiv_CTDT_DT}, we show that (\ref{eq:PAPR_CDF_DT_full}) closely approximates (\ref{eq:PAPR_CDF_CTDT}) when the support of $S_{ZZ}^d(f_d)$ is concentrated near $f_d=\lambda_1^d$.  

Note that (\ref{eq:PAPR_CDF_DT_full}) only depends on the spectral content of $Z_i$ through $\kappa = |r_{ZZ}[1]|$, which does not depend on the phase of $r_{ZZ}[k]$.  Therefore, (\ref{eq:PAPR_CDF_DT_full}) is invariant to spectral shifts of $Z_i$, which is consistent with the observation in Sec.~\ref{sec:foundations_PAPR} that the PAPR distribution does not change under frequency shifts of the baseband process.  

Since (\ref{eq:Psi_Q}) requires numerical evaluation of the Marcum-Q function, which can lead to inaccuracies in some circumstances, in the evaluations below, we calculated (\ref{eq:PAPR_CDF_DT_full}) with $\Psi(x,\kappa)$ specified by (\ref{eq:Psi_integral}).  To control numerical error, the combined function $[e^{-x}/(1-e^{-x})]\Psi(x,\kappa)$ was evaluated using adaptive quadrature as implemented by the Python SciPy \cite{SciPy} \texttt{integrate.quad} method with default parameters.  Because the integrand of (\ref{eq:Psi_integral}) has a singularity at $\theta = -\pi/2$ as $\kappa \rightarrow 1$, the \texttt{points} argument was used to break up the integration interval.

\subsection{A Refinement}
\label{sec:refinement}

\begin{figure}[tb]
    \centering
    \includegraphics[width = 0.75\columnwidth]{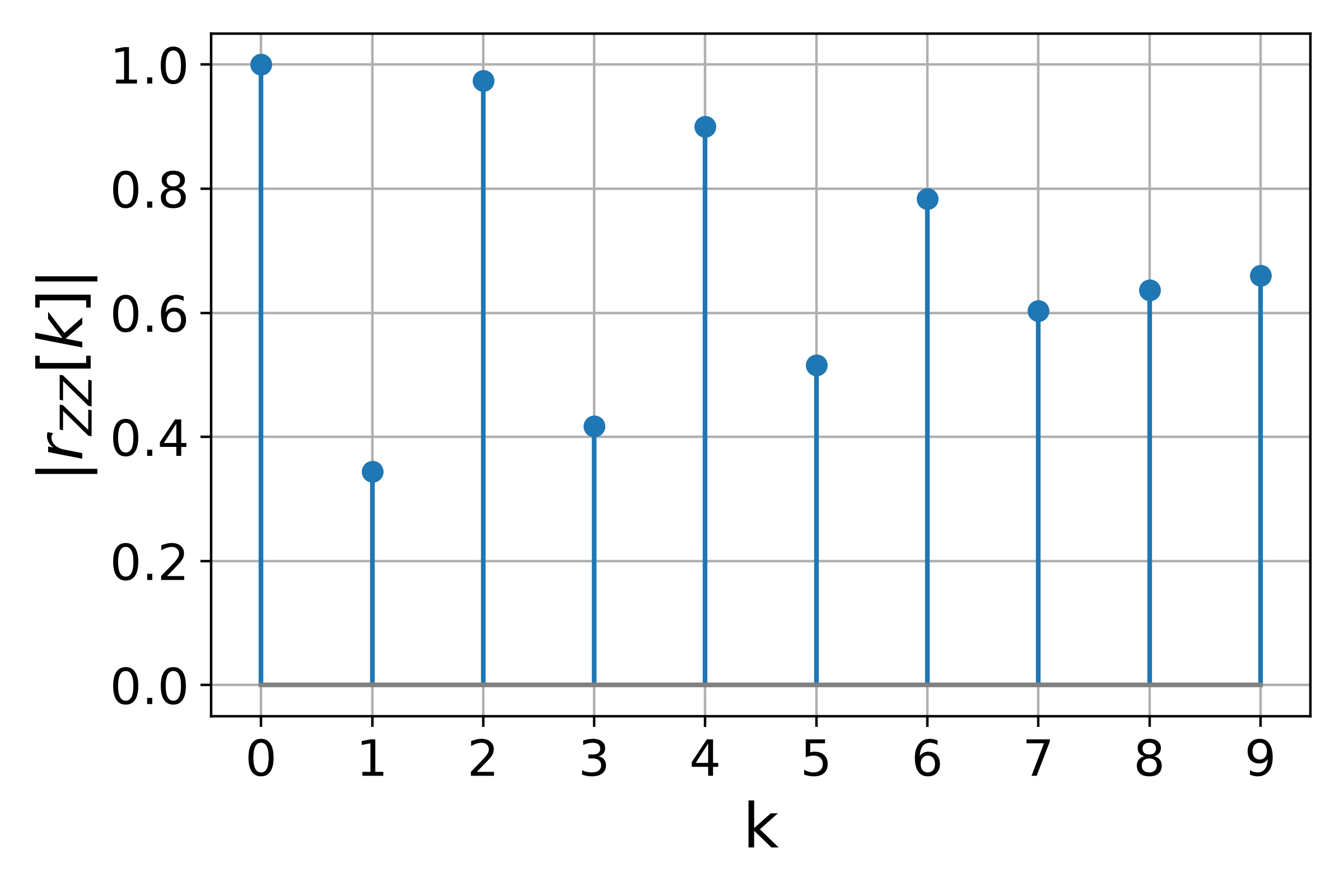}
    \caption{Example of a non-monotonic ACF magnitude. Corresponds to the $\mathrm{OSR}_N = 2$, $\mathrm{OSR}_L = 16$ case in Fig.~\ref{fig:asymmetric_2comp_BP}.}
    \label{fig:kappa_hat}
\end{figure}

When $\kappa = |r_{ZZ}[1]|$ is smaller than $|r_{ZZ}[k]|$ for some $k>1$, the first-order Markov model underlying (\ref{eq:PAPR_CDF_DT_full}) may be poor.  For example, consider the ACF magnitude shown in Fig.~\ref{fig:kappa_hat}.  In this example, $\kappa = |r_{ZZ}[1]| \approx 0.344$ and $|r_{ZZ}[2]| \approx 0.974$, and the first-order Markov assumption is clearly inadequate. 

To extend (\ref{eq:PAPR_CDF_DT_full}) to such cases, we replace the parameter $\kappa$ with $\widehat{\kappa}$, defined as the maximum value of the ACF magnitude over the first $L-1$ positive lags.  Namely, 
\begin{equation}
    \widehat{\kappa} = \max_{1 \leq k \leq L-1} |r_{ZZ}[k]|.
    \label{eq:kappa_hat}
\end{equation}
This model refinement has the attractive property of reducing to the prior definition of $\kappa$ when the maximum occurs at $k=1$.   

The empirical results in Sec.~\ref{sec:results} support our choice of $\widehat{\kappa}$; where we used $L=10$.  Specifically, the above example corresponds to the $\mathrm{OSR}_N = 2$, $\mathrm{OSR}_L = 16$ case in Fig.~\ref{fig:asymmetric_2comp_BP}, where it is seen that (\ref{eq:PAPR_CDF_DT_full}) with $\widehat{\kappa}$, denoted ``DT'', outperforms (\ref{eq:PAPR_iid}) and (\ref{eq:PAPR_CDF_CTDT}).  Theoretical justification for our choice of $\widehat{\kappa}$ remains an open question that is outside the scope of the present study.  

\section{Evaluation Methods}
\label{sec:eval_methods}

\subsection{Filtered Complex White Gaussian Noise}

\begin{figure}[tb]
    \centering
    \includegraphics[width = \columnwidth]{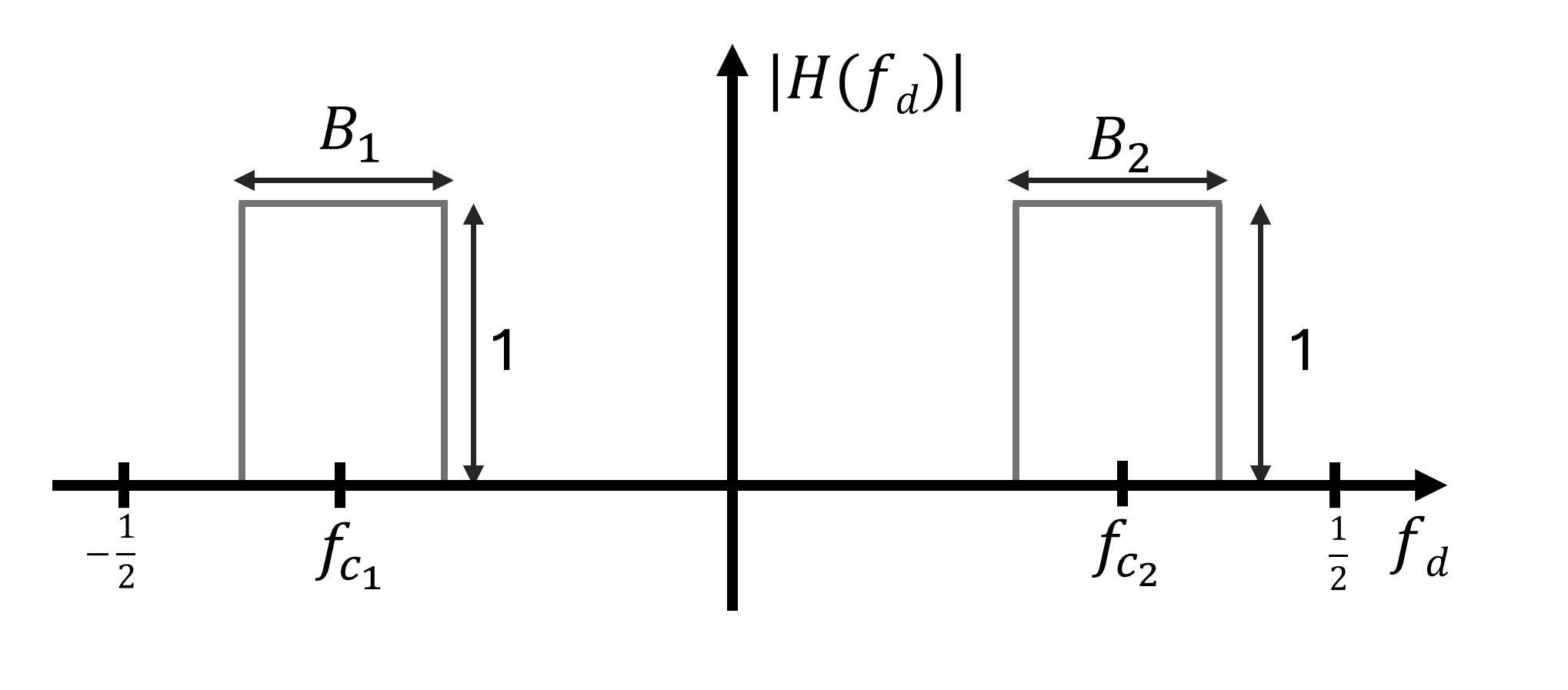}
    \caption{Parameters for band-pass filter with two pass-bands.}
    \label{fig:BP_filter_diagram}
\end{figure}

The effectiveness of different models for the PAPR distribution was evaluated numerically with Monte Carlo simulations of filtered CWGN. The general form of the ideal filter frequency response magnitude is depicted in Fig.~\ref{fig:BP_filter_diagram}, with two non-overlapping pass-bands, centered at $f_{c_1}$ and $f_{c_2}$, with $f_{c_1} < f_{c_2}$ and bandwidths $B_1$ and $B_2$. The filter parameters were expressed in normalized digital frequency units of cycles/sample so that $f_{c_1}, f_{c_2} \in [-1/2, 1/2)$, and $B_1, B_2 \in [0,1/2]$. 

The general dual-band filter frequency response was selected because it flexibly incorporates many cases of interest, including single-band and dual-band scenarios that may be asymmetric with respect to zero frequency.  Furthermore, this formulation enables straightforward calculation of the ACF magnitude and spectral dispersion for the filtered process, which are required to evaluate (\ref{eq:PAPR_CDF_CTDT}) and (\ref{eq:PAPR_CDF_DT_full}).  In particular, normalizing $S_{ZZ}^d(f_d)$ so that $\lambda_0 = 1$, the corresponding discrete-time ACF is 
\begin{align}
    r_{ZZ}[k] = \frac{1}{(B_1 + B_2)}\bigg(&B_1 \mathrm{sinc}(B_1 k)e^{j2\pi f_{c_1}k} \nonumber \\ &+ B_2 \mathrm{sinc}(B_2 k)e^{j 2\pi f_{c_2}k}\bigg) 
\end{align}
where $\mathrm{sinc}(x) = \sin(\pi x)/(\pi x)$ is the normalized sinc function. The ACF magnitude is easily seen to be 
\begin{align}
        |r_{ZZ}[k]| = \frac{1}{(B_1 + B_2)}&\bigg|B_1 \mathrm{sinc}(B_1 k)\nonumber \\ &+ B_2 \mathrm{sinc}(B_2 k)e^{j 2\pi (f_{c_2}-f_{c_1})k}\bigg|. \label{eq:ACF_mag}
\end{align}
Elementary, but tedious, calculations yield the following expression for the spectral dispersion
\begin{align}
    \lambda^d_2 - (\lambda^d_1)^2 = \frac{\pi^2}{3}(B_1^2 &- B_1B_2 + B_2^2) \nonumber \\&+ \frac{4\pi^2B_1B_2}{(B_1 + B_2)^2}\left(f_{c_2} - f_{c_1}\right)^2. \label{eq:spec_var}
\end{align}
Observe that (\ref{eq:ACF_mag}) and (\ref{eq:spec_var}) depend on the center frequencies only through the difference, $f_{c_2}-f_{c_1}$, which is consistent with the expectation of frequency shift invariance.   

The ideal filter response was approximated with a digital infinite impulse response (IIR) Butterworth filter of order $64$ implemented with second-order sections (SOS). Because practical digital filters require finite roll-offs, the effective bandwidths in our evaluations are slightly larger than the nominal values.  If the target PSD was symmetric with respect to zero frequency (i.e., even), then a single real-valued filter was used.  By contrast, if the target PSD was asymmetric around zero frequency, a complex-valued BP filter was used for each pass-band, and the filtered noise sequence was calculated as the sum of the filter outputs.  Complex BP SOS filter coefficients were synthesized from real-valued LP SOS filter coefficients using the method of Crystal and Ehrman \cite{Crystal1968}, which applies a suitable phase-shift to translate the pass-band to the desired center frequency.  

Each realization of the complex Gaussian process of length $n$ was synthesized as follows.  First, two independent sequences of i.i.d. standard normal deviates of length $n+8192$ were generated to form the real and imaginary components for a realization of CWGN with zero mean and variance two.  Next, digital filtering was applied to the CWGN sequence to obtain a length $n+8192$ complex filtered noise sequence.  Last, the middle $n$ samples were extracted, with the remaining samples discarded to avoid filtering transients.  Each Monte Carlo trial, consisting of one realization of the length $n$ filtered noise process, resulted in a single PAPR value, calculated as the ratio of the maximum to the mean.  

For each combination of filter and $n$, a total of $10^5$ independent Monte Carlo trials were executed, resulting in $10^5$ independent realizations of PAPR. The empirical CDF defined by the Monte Carlo samples was compared to the CDFs in (\ref{eq:PAPR_iid_gumbel}), (\ref{eq:PAPR_CDF_CTDT}) and (\ref{eq:PAPR_CDF_DT_full}) (with $\kappa$ replaced by $\widehat{\kappa}$ for $L=10$), corresponding to the i.i.d., continuous-time, and discrete-time models, respectively.  

\subsection{Experimental Parameterization}

As discussed in the Sec.~\ref{sec:introduction}, the degree of oversampling is expected to influence how the i.i.d. and continuous-time PAPR models perform.  Therefore, to facilitate comparisons, we parameterized numerical experiments in terms of two types of OSR that arise from classical sampling theory and provide complementary descriptions for the spectral support for the baseband process.  The first OSR is referenced to the Nyquist sampling rate \cite{Oppenheim1999}, which is sufficient but not necessary to avoid aliasing.  The second OSR is referenced to the theoretical minimum Landau rate \cite{Landau1967, Mishali2009}, which is necessary but not sufficient to avoid aliasing for 1-D uniform sampling.

Before defining the OSRs, we require some notation.  Let $\Omega = \operatorname{supp}(S_{ZZ}^d)$ denote the spectral support of the baseband process and let $\mathcal{I}$ be the closed convex hull of $\Omega$.  Assuming that the spectral support does not wrap around $f_d = \pm 1/2$, the convex hull is an interval of the form $\mathcal{I} = [f_{\mathrm{min}},f_{\mathrm{max}}]$, where $f_{\mathrm{min}} = \min(\Omega)$ and $f_{\mathrm{max}} = \max(\Omega)$.  Specifically, consider the filtered CWGN process resulting from the filter in Fig.~\ref{fig:BP_filter_diagram}.  In the single-band case with $B_1 \neq 0, B_2 = 0$, $\mathcal{I} = [f_{c_1}-B_1/2, f_{c_1}+B_1/2]$ and in the dual-band case with $B_1 \neq 0, B_2 \neq 0$, $\mathcal{I} = [f_{c_1}-B_1/2, f_{c_2}+B_2/2]$.  

Denoting Lebesgue measure by $\mu(\cdot)$, define the normalized discrete-time Nyquist rate\footnote{Although the classical Nyquist rate is defined for low-pass spectra \cite{Oppenheim1999}, we use this generalized definition since the PAPR distribution is invariant to frequency shifts of $Z_i$; see Sec.~\ref{sec:foundations_PAPR}.} as $f_N = \mu(\mathcal{I})$ and the discrete-time Landau rate as $f_L = \mu(\Omega)$.  For the PSD implied by the filter response in Fig.~\ref{fig:BP_filter_diagram} in the single-band case with $B_1 \neq 0, B_2 = 0$, $f_N = f_L = B_1$.  In the dual-base case with $B_1 \neq 0, B_2 \neq 0$, $f_N = \Delta f_c + (B_1+B_2)/2$, where $\Delta f_c = f_{c_2} - f_{c_1}$, and $f_L = B_1 + B_2$.  

Because we are working with PSDs defined in terms of normalized digital frequency, $f_d \in [-1/2,1/2)$, it follows that $f_N, f_L \in (0, 1]$.  Hence, we define the Nyquist and and Landau OSRs as $\mathrm{OSR}_N = 1/f_N$ and  $\mathrm{OSR}_L = 1/f_L$.  Since $f_L \leq f_N$, it follows that $\mathrm{OSR}_L \geq \mathrm{OSR}_N$. 

We parameterized our evaluations in terms of $\mathrm{OSR}_N$ and $\mathrm{OSR}_L$ by choosing the filter parameters shown in Fig.~\ref{fig:BP_filter_diagram} accordingly.  Specifically, the bandwidths were selected to satisfy $\mathrm{OSR}_L = 1/(B_1 + B_2)$, and in the dual-band case, the difference of center frequencies, $\Delta f_c$, was selected so that $\Delta f_c = 1/\mathrm{OSR}_N - (B_1+B_2)/2$.  

\subsection{Performance Assessment}
To assess the distance between the empirical PAPR distribution and the distribution given by each analytical model, we used the one-dimensional Wasserstein-1 ($W_1$) distance (or earth mover's distance) \cite{Kolouri2017} 
\begin{equation}
W_1(F_1,F_2) = \int_{-\infty}^\infty |F_1(x) - F_2(x)|\,dx
\label{eq:W1_CDF}
\end{equation}
where $F_1(x)$ and $F_2(x)$ are two CDFs. This metric was selected because (i) it can be directly computed from CDFs, (ii) it accounts for differences in both shape and location, and (iii) it has the same units as the underlying random variable.  

Since PAPR is typically converted to decibels (dB), i.e.,  $\text{PAPR}_{\text{dB}} = 10\log_{10}(\text{PAPR})$, we calculated $W_1$ in dB by directly estimating the empirical CDF on a dB scale and transforming the analytical PAPR CDFs to a dB scale using the CDF transformation rule \cite[Thm.~2.1.3]{Casella2002}.  The $W_1$ distance was estimated by integrating (\ref{eq:W1_CDF}) with adaptive numerical quadrature implemented by the Python SciPy \texttt{integrate.quad} method \cite{SciPy}.  Because each empirical distribution was estimated from $10^5$ independent Monte Carlo trials, relative uncertainties in the $W_1$ distance were negligible and uncertainty intervals were not plotted.    

Selected PDFs were also compared graphically.  PDFs for the empirical PAPR distribution were estimated using the kernel density method with a Gaussian kernel as implemented by \texttt{stats.gaussian\_kde} in Python SciPy with default settings \cite{SciPy}.  PDFs corresponding to (\ref{eq:PAPR_iid_gumbel}) and (\ref{eq:PAPR_CDF_CTDT}), which take the form of Gumbel distributions, were calculated from the Gumbel PDF transformed to a dB scale using the PDF transformation rule \cite[Thm.~2.1.5]{Casella2002}.  The PDF corresponding to (\ref{eq:PAPR_CDF_DT_full}), which is not a Gumbel form, was calculated numerically by first-order differencing the dB-scale CDF.  

\section{Monte Carlo Simulation Results}
\label{sec:results}

We assessed discrete-time, filtered CWGN in three scenarios: (A) symmetric single-band, (B) symmetric dual-band and (C) asymmetric dual-band.  Here, ``symmetric'' and ``asymmetric'' indicate whether the spectrum was symmetric with respect to zero frequency, i.e., if $\lambda_1^d = 0$.    

As explained in Sec.~\ref{sec:foundations_PAPR}, the PAPR distribution is invariant to frequency shifts, modulo 1, of $S_{ZZ}^d(f_d)$.  The models (\ref{eq:PAPR_iid_gumbel}) and (\ref{eq:PAPR_CDF_DT_full}) are likewise invariant to frequency translation, modulo 1, of $S_{ZZ}^d(f_d)$.  Similarly, as previously noted, (\ref{eq:PAPR_CDF_CTDT}) is invariant to frequency translation, provided the support of the shifted PSD is not split across the Nyquist boundary, $f_d= \pm 1/2$. Therefore, all empirical evaluations shown here do not change if the PSD $S_{ZZ}^d(f_d)$ is shifted, as long as the support of the shifted PSD is not split across the Nyquist boundary.  This theoretical expectation has been verified with additional simulations not presented here for brevity.  

In the plots and discussion below, the analytical models specified by (\ref{eq:PAPR_iid_gumbel}), (\ref{eq:PAPR_CDF_CTDT}), and (\ref{eq:PAPR_CDF_DT_full}) are denoted i.i.d, CT, and DT, respectively. Python code used to perform the evaluations is publicly available \cite{cg-papr}.

\subsection{Symmetric Single-Band}
For the symmetric single-band scenario, we set $B_1 = 2B$, $B_2=0$, and  $f_{c_1} = 0$, where the half-bandwidth was $B = 1/(2\,\mathrm{OSR}_L)$. In this case, $\mathrm{OSR}_N = \mathrm{OSR}_L$, the ACF magnitude is $|r_{ZZ}[k]| = |\mathrm{sinc}(2Bk)|$ and the spectral dispersion is $\lambda_2^d - (\lambda_1^d)^2 = 4\pi^2B^2/3$. 

\begin{figure}[t]
    \centering
    \subfloat[Filter frequency responses.\label{fig:symmetric_LP_filters}]{%
    \includegraphics[width=\columnwidth]{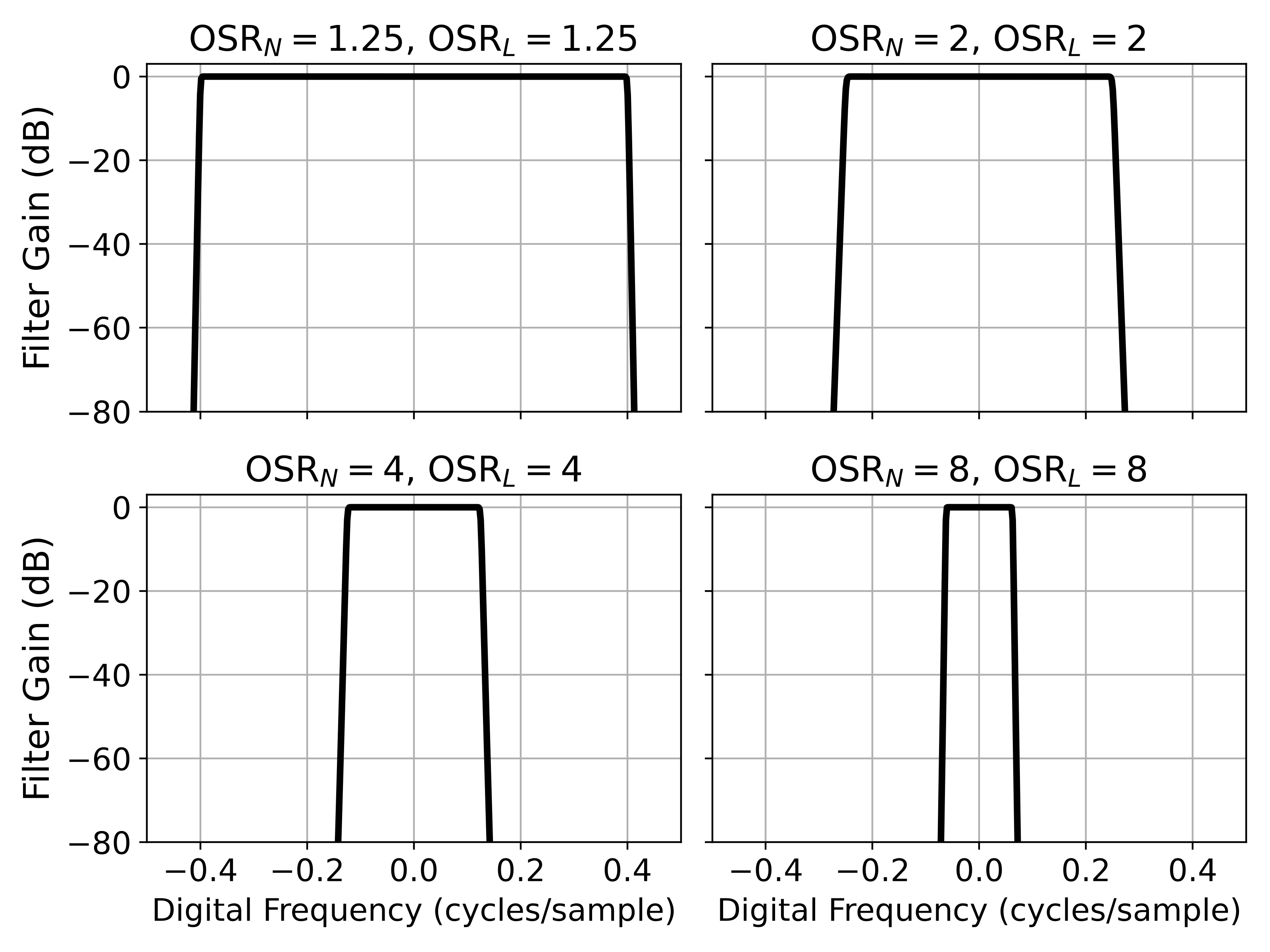}
    }\\
    \subfloat[Wasserstein-1 distance from empirical distribution.\label{fig:symmetric_LP_W1}]{%
    \includegraphics[width=\columnwidth]{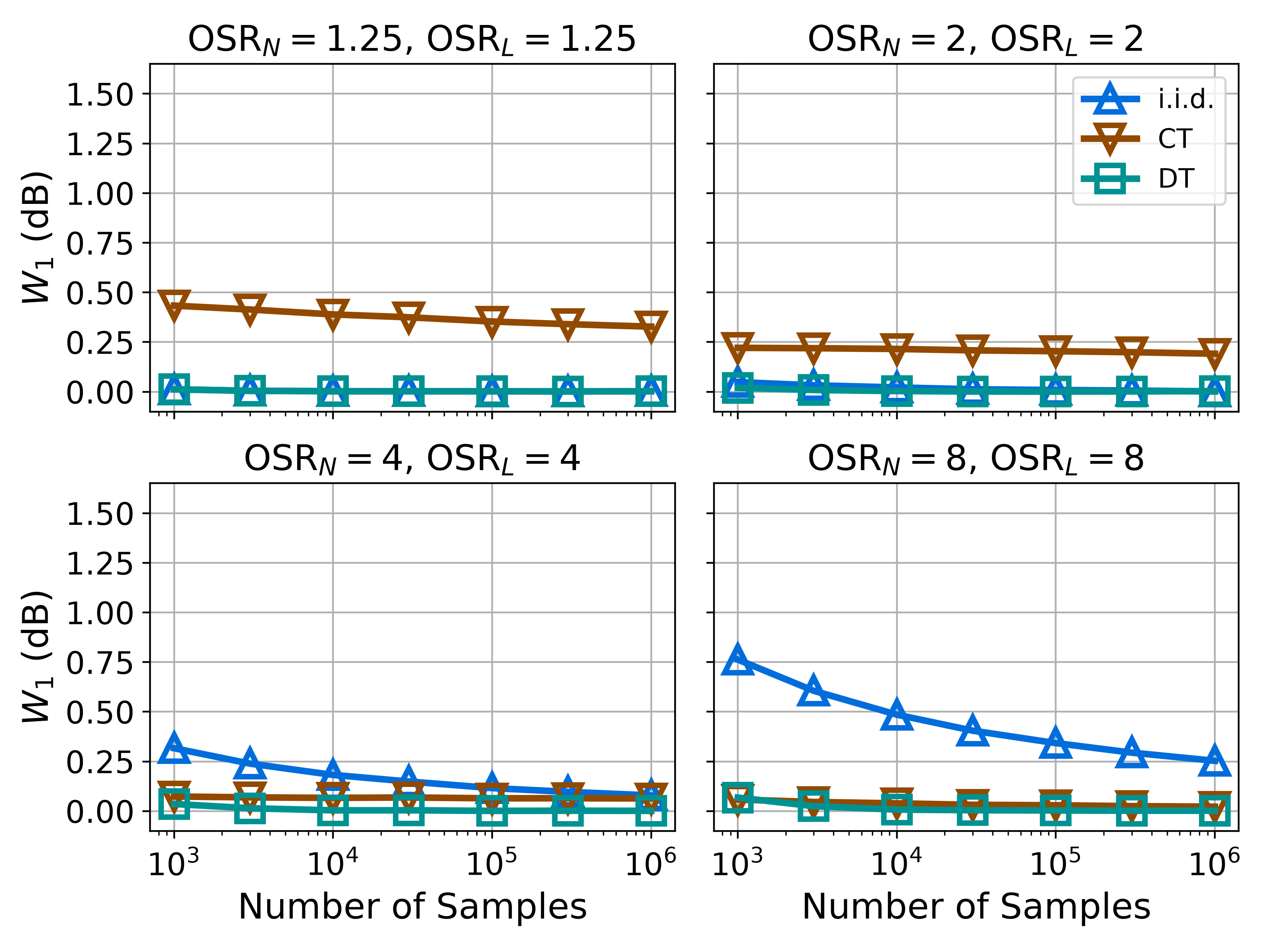}
    }\\
    \subfloat[Selected PDFs for $n=10\,000$. In the left plot, the i.i.d. model markers are not visible because they overlap with the empirical and DT model markers.  \label{fig:symmetric_LP_pdfs}]{%
    \includegraphics[width=\columnwidth]{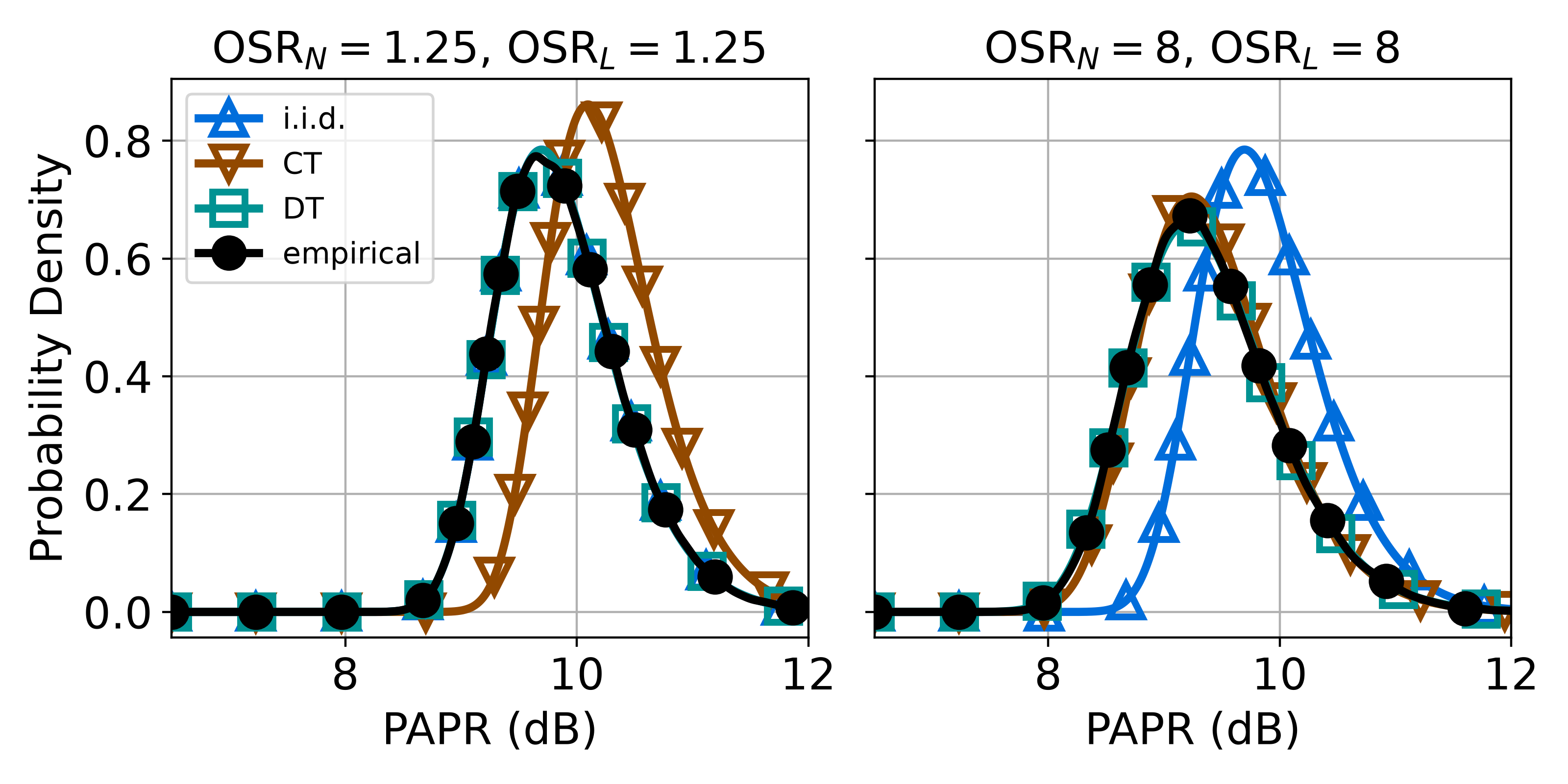}
    }
    \caption{Results for symmetric single-band filtered CWGN.}
    \label{fig:symmetric_LP}
\end{figure}

We assessed $\mathrm{OSR}_N = \mathrm{OSR}_L = 1.25, 2, 4, 8$, corresponding to $B = 0.4, 0.25, 0.125, 0.0625$ cycles/sample; see Fig.~\ref{fig:symmetric_LP_filters}.  Plots of $W_1$ distance versus sample $n$ are given in Fig.~\ref{fig:symmetric_LP_W1} and plots of selected PDFs for $n=10^4$ are shown in Fig.~\ref{fig:symmetric_LP_pdfs}.  

The results show that the empirical PAPR distribution is well approximated by the i.i.d. model for $\mathrm{OSR}_N \leq 2$, but that the i.i.d. model becomes progressively worse as $\mathrm{OSR}_N$ increases.  By contrast, the CT model closely approximates the empirical distribution for $\mathrm{OSR}_N \geq 4$.  In all cases, the new DT distribution model matches the empirical result very closely, indicating that it captures the regimes covered by both the i.i.d. model and the CT model, as anticipated.  

\subsection{Symmetric Dual-Band}
For the symmetric dual-band scenario, we set $B_1 = B_2 = B$ with $B = 1/(2\,\mathrm{OSR}_L)$ and set $f_{c_1} = -f_c$ and $f_{c_2} = f_c$ with $f_c = 1/(2\mathrm{OSR}_N) - B/2$.  In this case, the ACF magnitude is $|r_{ZZ}[k]| = |\mathrm{sinc}(Bk)\cos(\pi\Delta f_c k)|$ and the spectral dispersion is  $\lambda_2^d - (\lambda_1^d)^2 = \pi^2(B^2/3 + \Delta f_c^2)$.  

\begin{figure}[t]
    \centering
    \subfloat[Filter frequency responses.\label{fig:symmetric_BP_filters}]{\includegraphics[width=\columnwidth]{%
    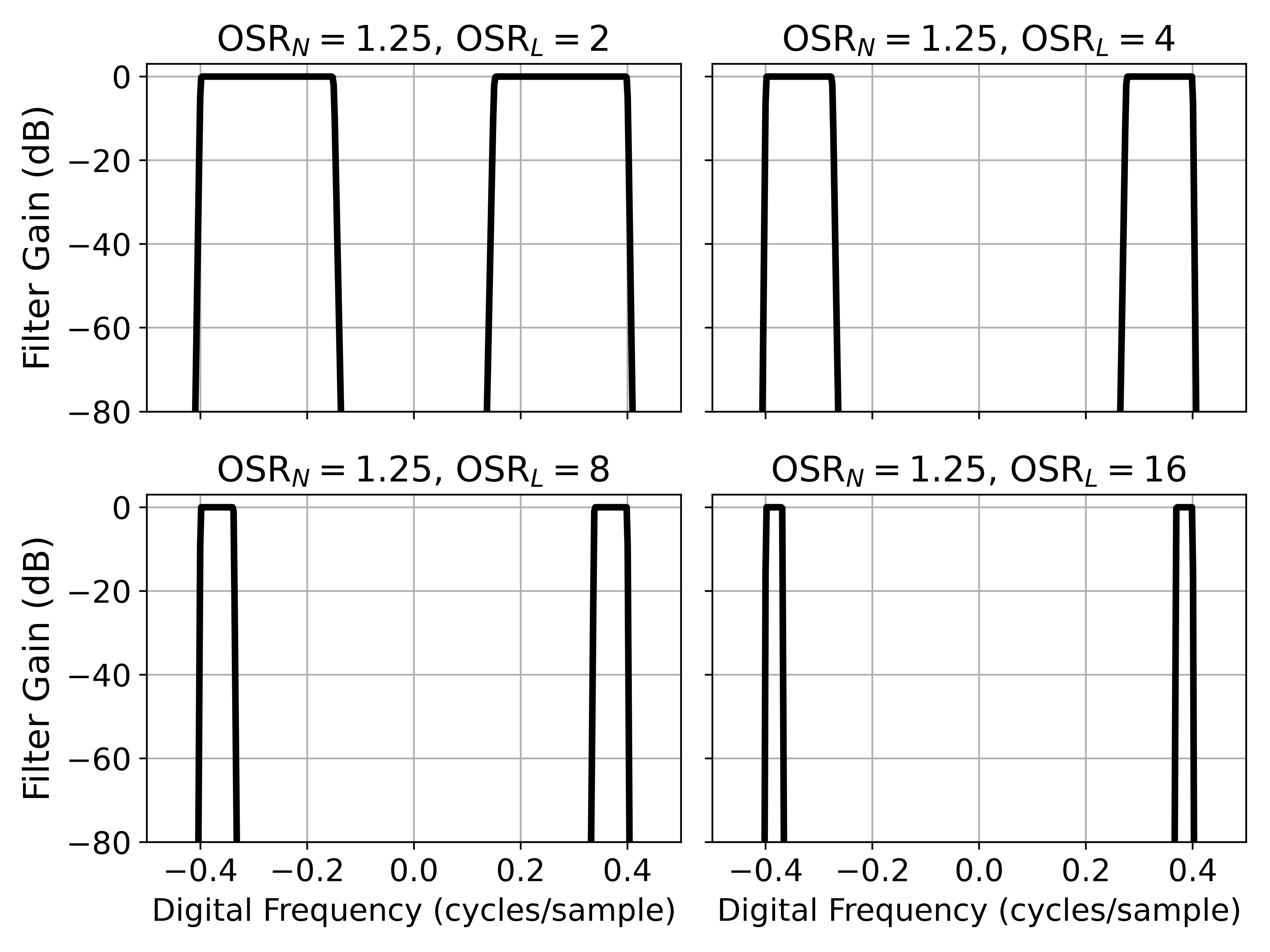}
    }\\
    \subfloat[Wasserstein-1 distance from empirical distribution.   \label{fig:symmetric_BP_W1}]{\includegraphics[width=\columnwidth]{%
    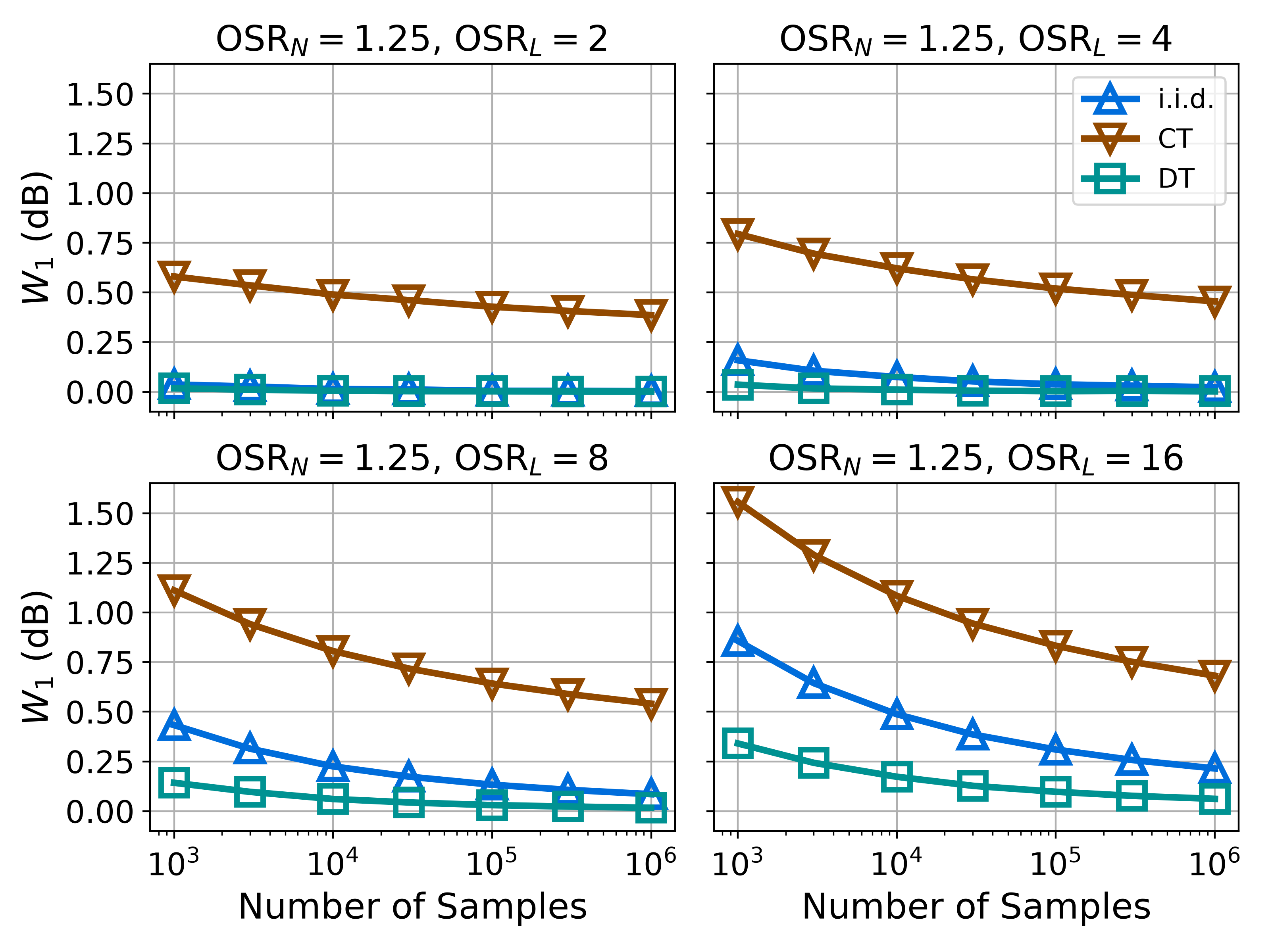}
    }\\
    \subfloat[Selected PDFs for $n=10\,000$. In the left plot, the i.i.d. model markers are not visible because they overlap with the empirical and DT model markers.   \label{fig:symmetric_BP_pdfs}]{%
    \includegraphics[width=\columnwidth]{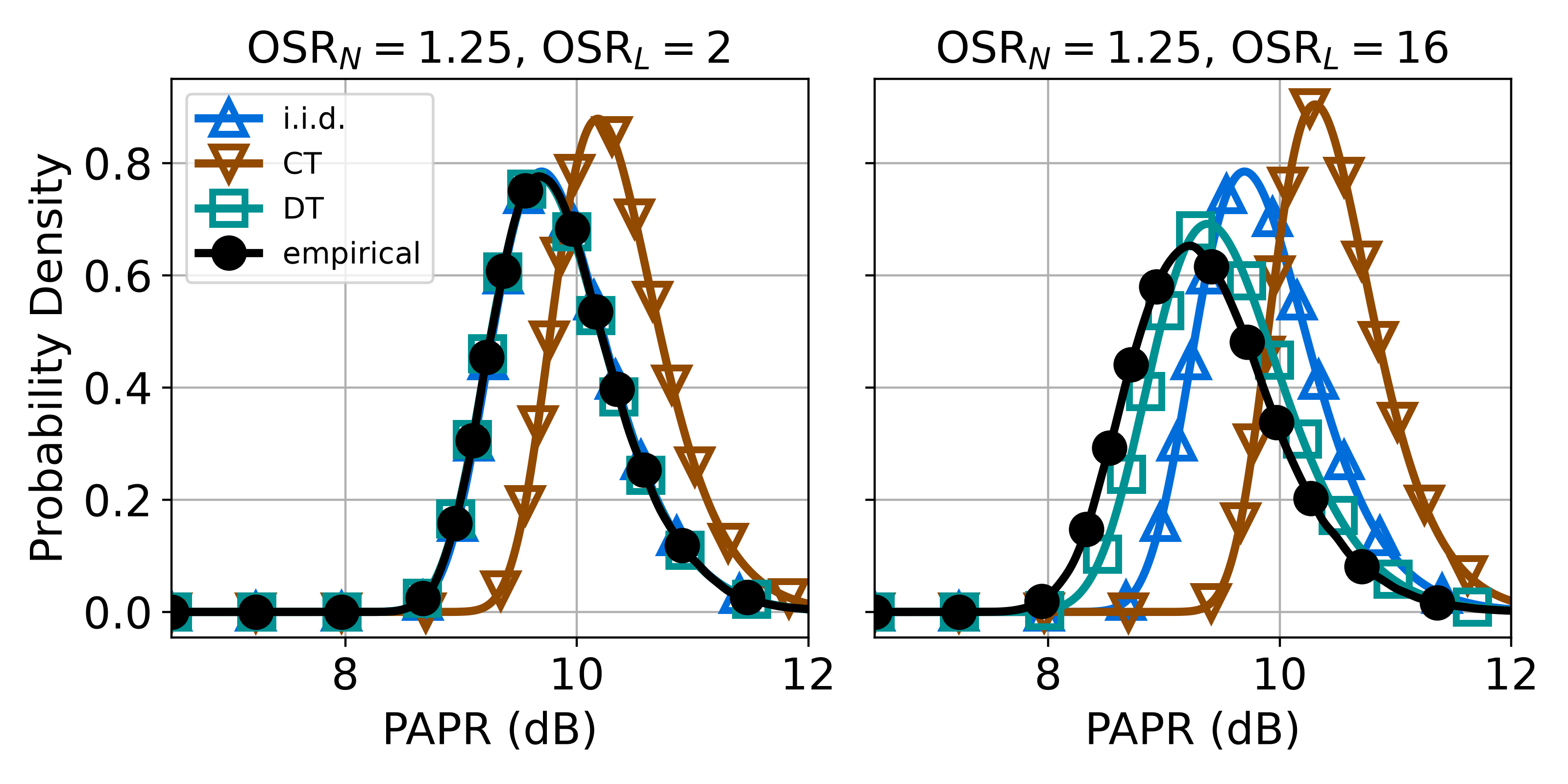}
    }
    \caption{Results for symmetric dual-band filtered CWGN with $\Delta f_c = 0.4$ cycles/sample.}
    \label{fig:symmetric_BP}
\end{figure}

We assessed $\mathrm{OSR}_N = 1.25$ with $\mathrm{OSR}_L = 2, 4, 8, 16$, corresponding to $B = 0.25, 0.125, 0.0625, 0.03125$ cycles/sample; see Fig.~\ref{fig:symmetric_BP_filters}. Plots of $W_1$ distance versus $n$ are given in Fig.~\ref{fig:symmetric_BP_W1} and plots of selected PDFs for $n=10^4$ are shown in Fig.~\ref{fig:symmetric_BP_pdfs}.  

In this dual-band scenario, the i.i.d. model was a good approximation for the empirical PAPR distribution for $\mathrm{OSR}_L = 2$, but became progressively worse for larger $\mathrm{OSR}_L$ values.  By contrast, in all cases, the CT model was worse than the i.i.d. model, with the $W_1$ distance from the empirical distribution increasing as $\mathrm{OSR}_L$ increased.  Again, the new DT model was superior or on par with the i.i.d. and CT models in all instances.   

\subsection{Asymmetric Dual-Band}
For the asymmetric dual-band scenario, we assessed $\mathrm{OSR}_L = 16$ with $\mathrm{OSR}_N = 1.25, 2, 4, 8$.  To make the scenario asymmetric, we used dissimilar bandwidths, $B_1=1/24\approx 0.042$ cycles/sample and $B_2 = 1/48 \approx 0.021$ cycles/sample.  Also, we set $f_{c_1} = -0.3$  cycles/sample and $f_{c_2} = f_{c_1} + \Delta f_c$ with $\Delta f_c = 1/\mathrm{OSR}_N - (B_1+B_2)/2$.  Fig.~\ref{fig:symmetric_BP_filters} shows the filter responses.  Expressions for the ACF magnitude and spectral dispersion are given by (\ref{eq:ACF_mag}) and (\ref{eq:spec_var}).

\begin{figure}[t]
    \centering
    \subfloat[Filter frequency responses.\label{fig:asymmetric_2comp_BP_filters}]{%
    \includegraphics[width=\columnwidth]{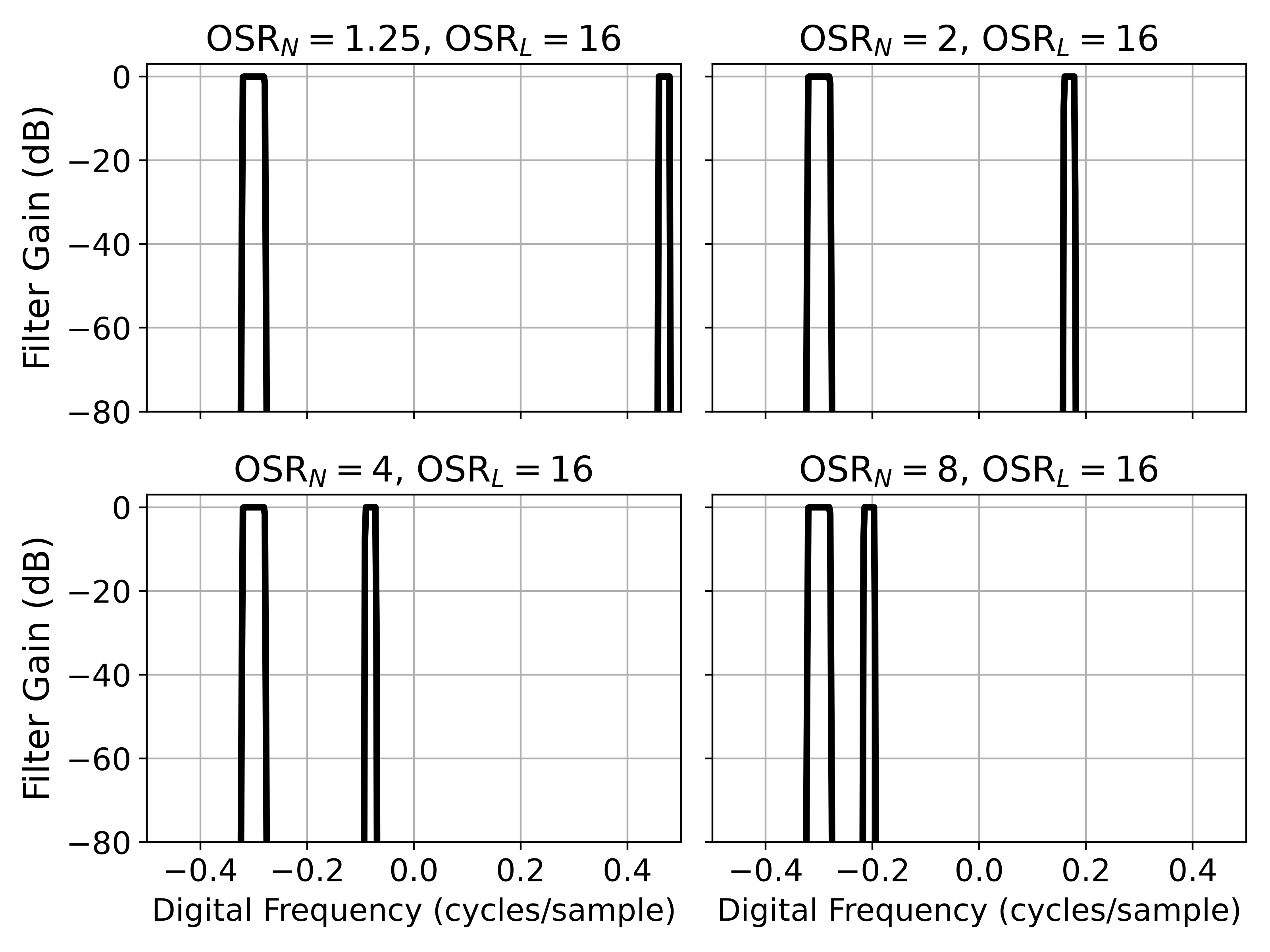}
    }\\
    \subfloat[Wasserstein-1 distance from empirical distribution.  \label{fig:asymmetric_2comp_BP_W1}]{%
    \includegraphics[width=\columnwidth]{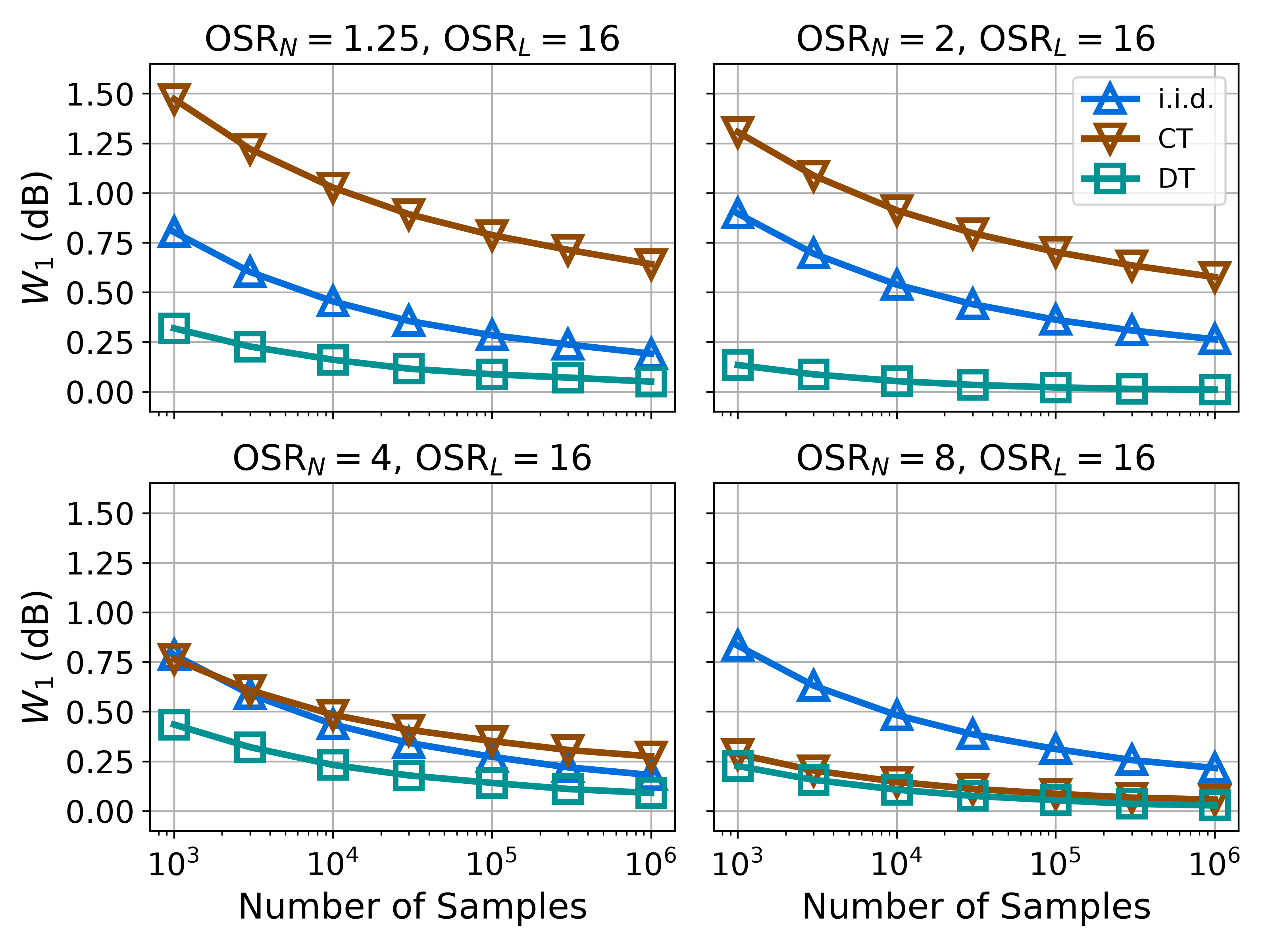}
    }\\
    \subfloat[Selected PDFs for $n=10\,000$.\label{fig:asymmetric_2comp_BP_pdfs}]{\includegraphics[width=\columnwidth]{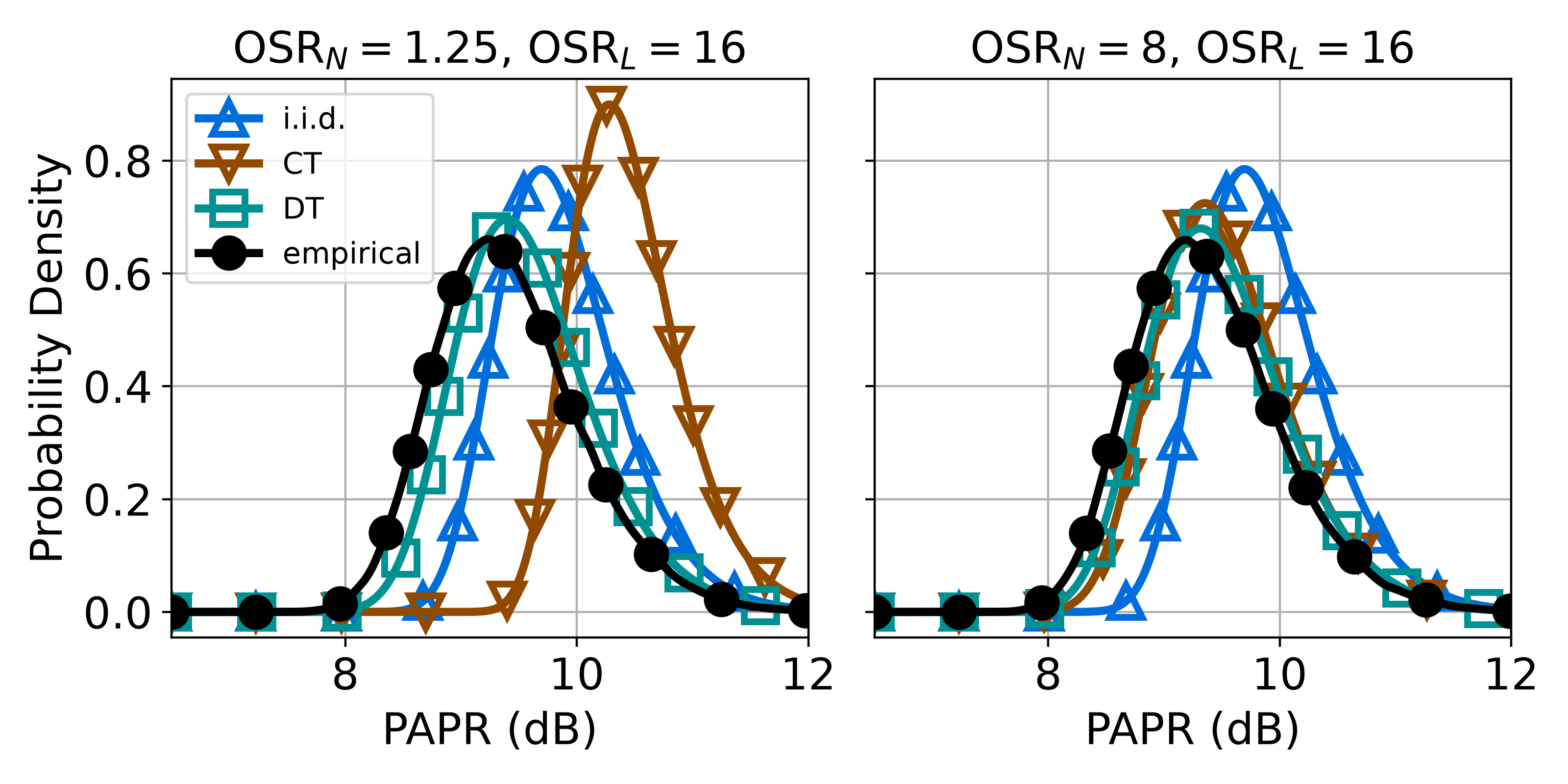}
    }
    \caption{Results for asymmetric dual-band filtered CWGN.} 
    \label{fig:asymmetric_2comp_BP}
\end{figure}

Plots of $W_1$ distance versus $n$ are given in Fig.~\ref{fig:asymmetric_2comp_BP_W1} and plots of selected PDFs for $n=10^4$ are shown in Fig.~\ref{fig:asymmetric_2comp_BP_pdfs}.  The i.i.d. model had similar discrepancies with the empirical distribution in all cases. By contrast, the accuracy of the CT model improved as $\mathrm{OSR}_N$ increased.  Like the prior scenarios, the DT model was superior or on par with the i.i.d. and CT models in all cases.  

\section{Discussion \& Conclusions}
\label{sec:conclusions}

We presented a new analytical model for the PAPR distribution of stationary, discrete-time complex Gaussian processes that arise from the baseband representation of band-limited, modulated random processes.  The new discrete-time model, which is based on a first-order, two-state Markov model for level exceedances, was demonstrated to bridge the established asymptotic i.i.d. and continuous-time models.  Moreover, the new discrete-time model was superior or on par with both the i.i.d. and continuous-time models in all scenarios examined.  

One way to potentially improve the discrete-time model would be to extend the first-order Markov model for level exceedances in Sec.~\ref{sec:markov_model} to a higher-order Markov chain.  A higher-order Markov model would require evaluating the joint exponential (or Rayleigh) CDF of three or more dependent variables to compute state transition probabilities.  Prior work on multivariate Rayleigh and exponential distributions, e.g., \cite{Isaac2019,Kotz2000}, may provide a starting point.

By linking our evaluations to $\mathrm{OSR}_L$ and $\mathrm{OSR}_N$, which characterize the relative size of the spectral support for $Z_i$, and its convex hull, respectively, we gained insights into each model's domain of efficacy. The i.i.d. model was accurate when $\mathrm{OSR}_L \leq 2$ and the continuous-time model was effective for $\mathrm{OSR}_N \geq 4$.  This latter finding is consistent with prior work on low-pass, continuous-time OFDM signals \cite{Tellambura2001}.  Based on our limited assessments, it appears that the performance of the i.i.d. and discrete-time models is largely driven by $\mathrm{OSR}_L$ and that the performance of the continuous-time model is primarily determined by $\mathrm{OSR}_N$.  Verification and elucidation of this observation is a interesting topic for future work.  

Another contribution, given in Appendix~A, was a connection between the literature on extreme value theory for continuous-time random processes and engineering works studying PAPR for OFDM.  Specifically, apparently different expressions given in \cite{Ochiai2001,Wei2002,Wei2003_thesis,Jiang2008} for the continuous-time OFDM PAPR distribution were shown to be asymptotically equivalent.   

There is a substantial body of work on asymptotic extreme value theory and PAPR for the envelope of band-limited, continuous-time Gaussian processes.  However, much less is known about the corresponding problem for discrete-time, sampled processes, which is pertinent to practical measurements with modern digital instrumentation.  The present work sheds light on this overlooked topic by introducing a new analytical model for the PAPR distribution that can be used to better characterize peak power in RF signals.  

\section*{Acknowledgments}
The author thanks Dazhen Gu, Susanna Mosleh, and Jake Rezac for helpful comments on the manuscript.  Google Gemini\footnote{Commercial software is identified in this paper only to accurately describe the implementation. This identification does not imply NIST endorsement or recommendation, nor does it mean these products are necessarily the best available for the task.} was used to perform routine calculations leading to (\ref{eq:spec_var}); the derivation was independently verified by hand.  All other content was solely conceived and developed by the author, who bears full responsibility for the integrity of the work.

\appendices

\section{PAPR Distribution for Continuous-Time Stationary Complex Gaussian Baseband Processes}
\label{appendix:CT_PAPR_theory}
We review established extreme value theory for the continuous-time process, $P(t)$, introduced in Sec.~\ref{sec:foundations}.  Furthermore, we connect this theory to the engineering literature on PAPR modeling for continuous-time OFDM signals.    

\subsection{Continuous-Time Theory}
As explained in Sec.~\ref{sec:foundations_PAPR}, without loss of generality, we take $\lambda_0=1$ throughout the paper.  However, in this subsection, we explicitly track dependence on $\lambda_0$ to elucidate dimensional consistency and to enable direct comparisons to expressions in the references that use different normalizations.    

In a nutshell, classical theory \cite{Cramer1967,Leadbetter1983} for stationary continuous-time random processes often approximates the extreme value distribution by arguing that the point process of level upcrossings asymptotically converges to a Poisson point process.  In this case, upcrossings are independent events and the number of upcrossings across a given level is Poisson distributed with rate parameter equal to the mean number of upcrossings per unit time.  Thus, the CDF for the maximum over a given time interval can be approximated by evaluating the Poisson distribution for the event of no upcrossings.   

For our problem setting with generally nonzero $\lambda_1$, the formula for the mean upcrossing rate of the real envelope, $R(t)$, was derived by Cram{\'e}r \& Leadbetter in their seminal monograph \cite[Sec.~11.8]{Cramer1967}.\footnote{When $\lambda_1=0$, the level crossing rate formula is a special case of an earlier result of Rice \cite[Eq. (4.8)]{Rice1948}.} An alternative derivation can be found in \cite[p.~165]{Vanmarcke2010}.  From this formula, it follows that the mean number of upcrossings of level $x\geq 0$ by the squared envelope, $P(t) = R^2(t)$, per unit time is\footnote{Cram{\'e}r \& Leadbetter \cite{Cramer1967} define spectral moments in terms of the PSD for the real passband process, unlike our definition in terms of the complex baseband process. Consequently, because the variance (or power) of the complex baseband process is twice that for the real passband representation \cite[Eq. (4.6.6)]{Proakis2002}, the formula in \cite{Cramer1967} includes additional factors of two.} 
\begin{equation}
    \nu_{P,x}^+ = e^{-x/\lambda_0}\delta\sqrt{\frac{x\lambda_2}{\pi\lambda_0^2}},
    \label{eq:CT_LCR}
\end{equation}
where $\delta = \sqrt{1 - \lambda_1^2/(\lambda_0\lambda_2)}$ is a dimensionless spectral width parameter satisfying $0 \leq \delta \leq 1$ \cite[p.~146]{Vanmarcke2010}. 

Sufficient conditions for convergence of level upcrossings for $P(t)$ to a Poisson point process with parameter $\nu_{P,x}^+T \rightarrow \Lambda(x) > 0$ as $x, T \rightarrow \infty$ were given by Lindgren \cite[Remark~1.2, Thm~2.1]{Lindgren1980}.  Under our assumptions, the only conditions that are not immediately satisfied are the so-called ``Berman'' conditions $r_{II}(\tau)\log(\tau) \rightarrow 0$ and $r_{IQ}(\tau) \log \tau \rightarrow \infty$ as $\tau \rightarrow \infty$. By (\ref{eq:z_acf}), these conditions are equivalent to 
\begin{equation}
    r_{ZZ}(\tau)\log(\tau) \rightarrow 0 \quad \text{as} \quad \tau \rightarrow \infty.
    \label{eq:generalized_berman}
\end{equation}
If $S_{ZZ}(f)$ is integrable with bounded variation, then $r_{ZZ}(\tau) = \mathcal{O}(1/|\tau|)$ \cite[Thm.~1]{Ghodadra2015}, and (\ref{eq:generalized_berman}) is satisfied.  Most processes of practical interest fulfill these requirements.  

When the number of upcrossings is Poisson distributed with parameter $\Lambda(x) \approx \nu_{P,x}^+T$, the CDF for $M_T = \max_{[0,T]} P(t)$ is the probability of no upcrossings across level $x$,
\begin{equation}
    F_{M_T}(x) = P(M_T \leq x) \approx \exp(-\nu_{P,x}^+T).
    \label{eq:MT_CT_Poisson}
\end{equation}
Because (\ref{eq:CT_LCR}) has a global maximum at $x=\lambda_0/2$, (\ref{eq:MT_CT_Poisson}) is not a valid CDF over the full range of $x \in [0,\infty)$.  If it is required to evaluate the CDF for $x \in [0, \lambda_0/2)$, this issue can be addressed by employing corrections first introduced by Vanmarcke in the context of real Gaussian processes \cite[p.~170]{Vanmarcke2010}\cite{Vanmarcke1975}.  In this work, we instead utilize an asymptotically equivalent form of the CDF that is presented next.

The extremal types theorem for stationary random processes \cite[Sec.~13.1]{Leadbetter1983} \cite[Sec.~3.1]{Leadbetter1988} implies that the extreme value distribution for $M_T$ is asymptotically equivalent to a Gumbel distribution. For this reason, and because the Gumbel distribution is a well-studied distribution with closed-form expressions for its PDF, quantiles, and moments, it is important to see how (\ref{eq:MT_CT_Poisson}) can be approximated by a Gumbel distribution.  

Specifically, following the approach of \cite{Sharpe1978, Lindgren1980,Leadbetter1988} for $\chi^2$-processes, suppose that $x$ and $T$ tend to infinity in a coordinated way, so that $(x/\lambda_0)/\log(T/T_0) \rightarrow 1$, where $T_0$ is a reference unit time base, e.g., $T_0 = 1\,ms$, included to keep the argument of the logarithm dimensionless \cite{Matta2011}\footnote{Although not used in \cite{Sharpe1978, Lindgren1980,Leadbetter1988}, inclusion of $\lambda_0$ and $T_0$ are necessary for dimensional consistency.}.  This condition implies that asymptotically,
\begin{equation}
    \log(x/\lambda_0) \approx \log\log(T/T_0). 
    \label{eq:nuT_inversion_approx}
\end{equation}
Substituting (\ref{eq:CT_LCR}) into $\Lambda(x) \approx \nu_{P,x}^+T =  \nu_{P,x}^+T_0(T/T_0)$, taking logarithms, and applying (\ref{eq:nuT_inversion_approx}) gives
\begin{align}
    \log \Lambda(x) \approx &- \frac{x}{\lambda_0} + \log\left(\frac{T}{T_0}\right) + \frac{1}{2}\log\log\left(\frac{T}{T_0}\right) \nonumber \\ &+\log\left(T_0\delta\sqrt{\frac{\lambda_2}{\pi\lambda_0}}\right).
\end{align}
Thus, as $x, T\rightarrow \infty$, (\ref{eq:MT_CT_Poisson}) can be approximated as
\begin{subequations}
\begin{align}
&F_{M_T}(x) \approx \exp[-\exp[-a_T(x-b_T)]] \\
&a_T = 1/\lambda_0 \\
&b_T = \lambda_0\Bigg[\log\left(\frac{T}{T_0}\right) + \frac{1}{2}\log\log\left(\frac{T}{T_0}\right)  \\ 
&\qquad +\log\left(T_0\delta \sqrt{\frac{\lambda_2}{\pi\lambda_0}}\right)\Bigg]. \nonumber
\end{align}
\label{eq:Gumbel_form}
\end{subequations}
Expressions equivalent to (\ref{eq:Gumbel_form}) with $\lambda_1=0$ (i.e., $\delta=1$) were given in \cite[p.~471]{Leadbetter1988} for $d$-degree of freedom $\chi^2$-processes.  

Now, normalizing $Z(t)$ so that $\lambda_0=1$, (\ref{eq:CT_LCR}) and (\ref{eq:MT_CT_Poisson}) imply that the CDF for PAPR is, without loss of generality, 
\begin{equation}
    F_{\text{PAPR}}(x) 
    \approx \exp\left[-Te^{-x}\sqrt{\frac{x}{\pi}(\lambda_2 - \lambda_1^2)}~\right].
    \label{eq:PAPR_CDF_CT_1}
\end{equation}
Similarly, (\ref{eq:Gumbel_form}) yields
\begin{equation}
    F_{\text{PAPR}}(x) \approx  \exp\left[-Te^{-x}\sqrt{\frac{1}{\pi}(\lambda_2 - \lambda_1^2)\log\left(\frac{T}{T_0}\right)}~\right].
    \label{eq:PAPR_CDF_CT_2}
\end{equation}
A formula equivalent to (\ref{eq:PAPR_CDF_CT_2}) without $T_0$ was stated by Wei \cite[Eq. (3.63)]{Wei2003_thesis}.  

Because the spectral dispersion, $\lambda_2 - \lambda_1^2$, does not change when $S_{ZZ}(f)$ is shifted in frequency, (\ref{eq:PAPR_CDF_CT_1}) and (\ref{eq:PAPR_CDF_CT_2}) are invariant to spectral shifts of the baseband process.  This is consistent with the frequency translation invariance of the continuous-time PAPR distribution discussed in Sec.~\ref{sec:foundations_PAPR}.

\subsection{Connections to Literature on the PAPR Distribution for Continuous-Time OFDM}

Now, we connect the above theory to the engineering literature on PAPR distributions for continuous-time OFDM signals \cite{Ochiai2001,Wei2002,Wei2003_thesis,Jiang2008}.  These works rely on the assumption that the complex baseband representation of a transmitted OFDM signal asymptotically converges to a complex Gaussian process as the number of subcarriers approaches infinity.  Wei et al. \cite{Wei2010} give results supporting the validity of this assumption when the observation time interval, $T$, is finite.        

Let $T = T_{\text{sym}}$ be the OFDM symbol period, $N$ be the number of subcarriers, and let the reference unit time base be $T_0 = T_{\text{sym}}/N$, so that $T = NT_0$.  Then (\ref{eq:PAPR_CDF_CT_1}) gives
\begin{equation}
    F_{\text{PAPR}}(x) 
    \approx \exp\left[-NT_0e^{-x}\sqrt{\frac{x(\lambda_2 - \lambda_1^2)}{\pi}}~\right],
    \label{eq:PAPR_OFDM_1}
\end{equation}
and (\ref{eq:PAPR_CDF_CT_2}) yields
\begin{equation}
    F_{\text{PAPR}}(x) \approx  \exp\left[-NT_0e^{-x}\sqrt{\frac{\lambda_2 - \lambda_1^2}{\pi}\log N}~\right].
    \label{eq:PAPR_OFDM_2}
\end{equation}
The above formulas, with the absence of the reference unit time base, correspond to expressions given by Wei \cite{Wei2003_thesis} and Jiang et al. \cite{Jiang2008} when the subcarrier power distribution is nonuniform.  Specifically, (\ref{eq:PAPR_OFDM_2}) is equivalent to \cite[Eq. (3.64)]{Wei2003_thesis}.  When $\lambda_1=0$, (\ref{eq:PAPR_OFDM_1}) is equivalent to \cite[Eq. (22)]{Jiang2008}, which calculates $\lambda_2$ directly from the subcarrier power profile.

References \cite{Ochiai2001,Wei2002,Wei2003_thesis} consider the special case where OFDM subcarriers have a uniform power distribution, which implies that the baseband process, $Z(t)$, has an ideal, two-sided symmetric lowpass spectrum with half-bandwidth $B = N/(2T_{\text{sym}})$, i.e., $S_{ZZ}(f) = 1/(2B)$ if $|f| \leq B$ and  $S_{ZZ}(f) = 0$ otherwise.  In this case, $\lambda_2 = (4\pi^2/3)B^2 = (\pi^2/3)(N/T_{\text{sym}})^2 = \pi^2/(3T_0^2)$ and $\lambda_1 = 0$, so (\ref{eq:PAPR_OFDM_1}) becomes
\begin{equation}
    F_{\text{PAPR}}(x) \approx \exp\left(-Ne^{-x}\sqrt{\frac{\pi}{3}x}~\right),
    \label{eq:OFDM_PAPR_3}
\end{equation}
which is the asymptotic result of Ochiai and Imai \cite[Eq. (27)]{Ochiai2001}.  Jiang et al. extend (\ref{eq:OFDM_PAPR_3}) to the case where not all subcarriers are active \cite[Eq. (18)]{Jiang2008}.  
Alternatively, (\ref{eq:PAPR_OFDM_2}) gives the asymptotically equivalent form
\begin{equation}
    F_{\text{PAPR}}(x) \approx \exp\left[-Ne^{-x}\sqrt{\frac{\pi}{3}\log N}~\right],
\end{equation}
corresponding to the formula of Wei et al. \cite[Eq. (17)]{Wei2002} \cite[Eq. (3.47)]{Wei2003_thesis} for uniform power across subcarriers.  

\section{Bivariate Exponential CDF}
\label{appendix:bivariate_exponential}

We outline the derivation of expressions for the bivariate exponential CDF used in Sec.~\ref{sec:practical_DT_PAPR_model}.

As noted in Sec.~\ref{sec:foundations_PAPR}, when $\lambda_0=1$, $P_i=R_i^2$ and $P_{i+1}=R_{i+1}^2$, are standard exponential random variables with unit mean and unit variance.  Furthermore, $R_i$ and $R_{i+1}$ are Rayleigh distributed with scale parameter $\sqrt{1/2}$.  Below, we give expressions for the bivariate CDF of $P_i$ and $P_{i+1}$ that follow directly from established results for the bivariate Rayleigh CDF of $R_i$ and $R_{i+1}$ \cite[Appendix~A]{Schwartz1966} \cite[Sec.~6.1]{Simon2005}. The bivariate exponential distribution arising from the bivariate Rayleigh distribution is one of several bivariate exponential distributions that have been studied in the literature; see [Ch~47, Sec. 2]\cite{Kotz2000} for a review.  

First, we examine how the correlation coefficient, $\rho = \mathrm{Cov}(P_i,P_{i+1})$, is related to $\kappa = |\mathrm{Cov}(Z_i,Z_{i+1})|$.  Namely,
\begin{align}
    \rho &= \mathrm{E}[P_i P_{i+1}] - \mathrm{E}[P_i]\mathrm{E}[P_{i+1}] \\
    &= \mathrm{E}[Z_iZ_i^*Z_{i+1}Z_{i+1}^*] - 1. \label{eq:rho}
\end{align}
Because $Z_i$ is a zero-mean, stationary, proper complex Gaussian process, a theorem of Reed \cite{Reed1962} implies 
\begin{align}
    \mathrm{E}[Z_iZ_i^*Z_{i+1}Z_{i+1}^*] &= \mathrm{E}[Z_iZ_i^*]\mathrm{E}[Z_{i+1}Z_{i+1}^*] \\
    &\qquad+  \mathrm{E}[Z_iZ_{i+1}^*]\mathrm{E}[Z_{i+1}Z_{i}^*] \\
    &= r_{ZZ}[0]^2 + r_{ZZ}[1]r_{ZZ}^*[1] \\
    & = 1 + |r_{ZZ}[1]|^2, \label{eq:moment_decomp}
\end{align}
where $r_{ZZ}[k] = \mathrm{E}[Z_{i+k}Z_i^*] = \mathrm{Cov}(Z_{i+k},Z_i)$ is the discrete-time ACF.  Combining (\ref{eq:rho}) and (\ref{eq:moment_decomp}) thus yields
\begin{equation}
    \rho = |r_{ZZ}[1]|^2 = \kappa^2.
\end{equation}

The bivariate Rayleigh CDF for $(R_i,R_{i+1})$ can be stated in terms on the Marcum-Q function when $0 \leq \rho < 1$ \cite[Eq.~A-7-4]{Schwartz1966} \cite[Eq.~6.5]{Simon2005}.  Substituting $(P_i,P_{i+1}) = (R_i^2,R_{i+1}^2)$ and $\kappa = \sqrt{\rho}$ into the bivariate Rayleigh CDF in \cite[Eq.~6.5]{Simon2005} gives 
\begin{align}
    F_{P_i,P_{i+1}}(x,y) &= P(P_i \leq x, P_{i+1} \leq y) \nonumber \\
    &= 1- e^{-x}Q_1(d,\kappa c) \nonumber \\
    &\qquad -e^{-y}[1-Q_1(\kappa d, c)],
    \label{eq:joint_exp_CDF}
\end{align}
where $0 \leq \kappa < 1$, $c = \sqrt{2x/(1-\kappa^2)}$, $d = \sqrt{2y/(1-\kappa^2)}$, and $Q_1(a,b)$ is the first-order Marcum-Q function.  Oberve that when $x=y$, (\ref{eq:joint_exp_CDF}) becomes (\ref{eq:joint_P_CDF_xx}) with $\Psi(x,\kappa)$ defined by (\ref{eq:Psi_Q}).  

When $Z_i$ and $Z_{i+1}$ are independent, $\kappa=0$, and the identities $Q_1(a,0)=1$ and $Q_1(0,b) = \exp(-b^2/2)$ \cite[p.~97]{Simon2005}, imply that $F_{P_i,P_{i+1}}(x,y) = (1-e^{-x})(1-e^{-y})$.  Thus,  the joint CDF for independent $P_i$ and $P_{i+1}$ reduces to the product of the marginal CDFs for independent, standard exponential random variables, as expected.

Simon and Alouini \cite[Sec.~6.1]{Simon2005} provide alternative expressions for the bivariate Rayleigh CDF in terms of a single integral with finite limits and that do not require special function evaluations.  In particular, starting with \cite[Eq.~6.9-6.11]{Simon2005}, making the substitutions $(P_i,P_{i+1}) = (R_i^2,R_{i+1}^2)$ and $\kappa = \sqrt{\rho}$, and letting $x=y$, straightforward simplifications result in (\ref{eq:joint_P_CDF_xx}) with $\Psi(x,\kappa)$ given by (\ref{eq:Psi_integral}).

\section{Approximate Equivalence of Discrete-Time and Continuous-Time PAPR Distributions for Concentrated Spectral Support}
\label{appendix:equiv_CTDT_DT}
When the spectral support of $Z_i$ is concentrated near $f_d = \lambda_1^d$,
we show that (\ref{eq:PAPR_CDF_DT_full}) is approximately equivalent to (\ref{eq:PAPR_CDF_CTDT}) as $x\rightarrow \infty$ and $n\rightarrow \infty$.  As stated in Sec.~\ref{sec:foundations_PAPR}, we take $\lambda_0=1$ without loss of generality.  

First, we relate $\kappa = |r_{ZZ}[1]|$ to the spectral spread, $\lambda_2^d - (\lambda_1^d)^2$.  Since $r_{ZZ}[k]$ and $S_{ZZ}^k(f_d)$ are a Fourier transform pair, 
\begin{align}
    r_{ZZ}[1] &= \int_{-1/2}^{1/2} S_{ZZ}^d(f_d) e^{j 2\pi f_d}\,df_d \\
    &= e^{j 2\pi \lambda_1^d}  \int_{-1/2}^{1/2} S_{ZZ}^d(f_d) e^{j 2\pi (f_d - \lambda_1^d)}\,df_d. \label{eq:rzz1_expansion} 
\end{align}
Supposing that the support of $S_{ZZ}^d(f_d)$ is contained in a small neighborhood of $f_d = \lambda_1^d$, the second-order Maclaurin approximation for the exponential term yields
\begin{align}
    r_{ZZ}[1] &\approx e^{j 2\pi \lambda_1^d}  \int_{-1/2}^{1/2} S_{ZZ}^d(f_d) \Big[ 1 + j2\pi (f_d - \lambda_1^d) \nonumber \\
    &\qquad- \frac{1}{2} (2\pi (f_d - \lambda_1^d))^2 \Big]\,df_d. \\
    &\approx e^{j 2\pi \lambda_1^d} \left[1 - \frac{1}{2}\left[\lambda_2^d - (\lambda_1^d)^2\right]\right].
\end{align}
Thus, 
\begin{align}
    \kappa^2 \approx  \left(1 - \frac{1}{2}\left[\lambda_2^d - (\lambda_1^d)^2\right]\right)^2 
    \approx 1 - \left[\lambda_2^d-(\lambda_1^d)^2\right], \label{eq:kappa_approx}
\end{align}
where the quadratic term was dropped in the last approximation, since $\lambda_2^d - (\lambda_1^d)^2$ is small.

As noted in Sec.~\ref{sec:markov_model}, $p_{00}(x) = 1 - p_{01}(x)$.  So (\ref{eq:p_00}) gives
\begin{equation}
    p_{01}(x) = \left(\frac{e^{-x}}{1-e^{-x}}\right) \Psi(x,\kappa). \label{eq:p_01} 
\end{equation}
As $x \rightarrow \infty$, (\ref{eq:Psi_integral}) and (\ref{eq:p_01}) imply that $p_{01}(x)$ is small.  Hence, (\ref{eq:CDF_DT}) yields
\begin{align}
      F_{\text{PAPR}}^{\text{DT}}(x) &= (1-e^{-x}) [1-p_{01}(x)]^{n-1} \\
      &\approx \exp[-(n-1)p_{01}(x)] \label{eq:exp_approx} \quad \text{as } x\rightarrow \infty\\
      &\approx \exp[-n p_{01}(x)] \quad \text{as } x\rightarrow \infty, n\rightarrow \infty \label{eq:DT_Poisson_approx}
\end{align}
where (\ref{eq:exp_approx}) used $(1-\epsilon)^n \approx e^{-n \epsilon}$ with $\epsilon = p_{01}(x)$. 

Since $\lambda_2^d - (\lambda_1^d)^2$ is small, by (\ref{eq:kappa_approx}), $\kappa \approx 1$.  Thus, (\ref{eq:Psi_integral}) implies
\begin{equation}
    \Psi(x,\kappa) \approx 1 - \frac{1}{2\pi} \int_{-\pi}^{\pi} \exp\left[-\frac{x(1-\kappa^2)}{2 + 2\sin\theta}\right]\,d\theta. \label{eq:Psi_approx1}
\end{equation}
The above integral can be restated using Craig's representation of the complementary error function \cite[Eq.~10]{Craig1991} \cite[Sec.~4.1]{Simon2005}
\begin{equation}
    \mathrm{erfc}(z) = \frac{2}{\pi} \int_0^{\pi/2} \exp\left( -\frac{z^2}{\sin^2(\phi)}\right) d\phi \quad \text{for } z\geq 0. \label{eq:craig_rep}
\end{equation}

Namely, since $2 + 2\sin\theta = 4 \sin^2(\theta/2 + \pi/4)$, (\ref{eq:Psi_approx1}) becomes
\begin{equation}
    \Psi(x,\kappa) \approx 1 - \frac{1}{2\pi} \int_{-\pi}^{\pi} \exp\left[-\frac{x(1-\kappa^2)}{4 \sin^2(\theta/2 + \pi/4)}\right]\,d\theta. 
\end{equation}
Making the substitution $\phi = \theta/2 + \pi/4$, and noting that the resulting integrand is $\pi$-periodic and even around $\phi = \pi/2$, 
\begin{equation}
    \Psi(x,\kappa)
     \approx 1 - \frac{2}{\pi} \int_{0}^{\pi/2} \exp\left[-\frac{x(1-\kappa^2)}{ 4\sin^2(\phi)}\right]\,d\phi. \label{eq:Psi_approx2}
\end{equation}
Applying (\ref{eq:craig_rep}) gives
\begin{align}
     \Psi(x,\kappa) &\approx 1 - \mathrm{erfc}\left(\sqrt{x(1-\kappa^2)/4} \right) \\
     &\approx \mathrm{erf}\left(\sqrt{x(1-\kappa^2)/4} \right), \label{eq:Psi_approx3}
\end{align}
where $\mathrm{erf}(z)$ denotes the error function. The leading term of the Maclaurin series for the error function is $\mathrm{erf}(z) \approx 2z/\sqrt{\pi}$ \cite[Eq.~7.6.1]{NIST-DLMF}.  Since $\kappa \approx 1$, the argument of (\ref{eq:Psi_approx3}) is small, so
\begin{equation}
     \Psi(x,\kappa) \approx \sqrt{\frac{x(1-\kappa^2)}{\pi}} \approx \sqrt{\frac{x[\lambda_2^d - (\lambda_1^d)^2]}{\pi}}, \label{eq:Psi_approx4}
\end{equation}
where we applied (\ref{eq:kappa_approx}).  

Finally, combining (\ref{eq:p_01}), (\ref{eq:DT_Poisson_approx}), and (\ref{eq:Psi_approx4}), we see that as $x\rightarrow \infty, n\rightarrow \infty$,  
\begin{equation}
    F_{\text{PAPR}}^{\text{DT}}(x) \approx \exp\left[ -n e^{-x}\sqrt{\frac{x[\lambda_2^d - (\lambda_1^d)^2]}{\pi}}\right]. \label{eq:CDF_DT_approx_CT}
\end{equation}
In Appendix~\ref{appendix:CT_PAPR_theory}, we saw that the continuous-time expressions (\ref{eq:PAPR_CDF_CT_1}) and (\ref{eq:PAPR_CDF_CT_2}) for the PAPR distribution are asymptotically equivalent.  In particular, when the reference time base is equal to the sampling period, i.e., $T_0 = T_s$, and $n = T/T_s$, (\ref{eq:PAPR_CDF_CT_1}) yields (\ref{eq:CDF_DT_approx_CT}), and (\ref{eq:PAPR_CDF_CT_2}) yeilds (\ref{eq:PAPR_CDF_CTDT}).  So (\ref{eq:CDF_DT_approx_CT}) and (\ref{eq:PAPR_CDF_CTDT}) are asymptotically equivalent.  Thus, we conclude that when the support of $S_{ZZ}^d(f_d)$ is concentrated near $f_d = \lambda_1^d$, 
\begin{equation}
    F_{\text{PAPR}}^{\text{DT}}(x) \approx  F_{\text{PAPR}}^{\text{CT}}(x) \quad \text{as } x\rightarrow \infty, n \rightarrow \infty.
\end{equation}

\bibliographystyle{IEEEtran}
\bibliography{references}

\end{document}